\documentclass[sigchi]{acmart}

\setcopyright{none}

\copyrightyear{2019} 
\acmYear{2019} 
\setcopyright{acmcopyright}
\acmConference[CHI 2019]{CHI Conference on Human Factors in Computing Systems Proceedings}{May 4--9, 2019}{Glasgow, Scotland Uk}
\acmBooktitle{CHI Conference on Human Factors in Computing Systems Proceedings (CHI 2019), May 4--9, 2019, Glasgow, Scotland Uk}
\acmPrice{15.00}
\acmDOI{10.1145/3290605.3300874}
\acmISBN{978-1-4503-5970-2/19/05}

\usepackage{booktabs} 

\usepackage{color}  
\definecolor{olive}{rgb}{0.42, 0.56, 0.14}
\definecolor{brightlavender}{rgb}{0.75, 0.58, 0.89}
\newcommand{\ab}[1]{\textcolor{olive}{\Small{[AB: #1]}}}

\newcommand{\nb}[1]{\textcolor{brightlavender}{\Small{[NB: #1]}}}

\newcommand{\rmed}[1]{}

\newcommand{\stats}[2]{({\small\it{M:#1, SD:#2}})} 
\newcommand{\stat}[1]{({\small\it{#1}})} 
\newcommand{\statnp}[1]{{\small\it{#1}}} 

\newcommand{\Part}[1]{~(P#1)}
\newcommand{\subjective}{\textsuperscript{$\dagger$}}

\usepackage{tikz}
\usepackage{multirow}
\def\checkmark{\tikz\fill[scale=0.3](0,.35) -- (.25,0) -- (1,.7) -- (.25,.15) -- cycle;}
\newcommand{\quotes}[1]{``#1''}

\newcounter{rcounter}

\newcommand{\hrem}[1]{\textcolor{red}{#1}}
\renewcommand{\hrem}[1]{} 

\newcommand{\add}[1]{\textcolor{black}{#1}}

\newcommand{\addf}[1]{\textcolor{black}{#1}}

\newcommand{\fix}[1]{\textcolor{black}{#1}}

\newcommand{\tab}[1]{\hspace{.0067\textwidth}\rlap{#1}}

\usepackage[export]{adjustbox}

\begin{document}
\title{An Exploratory Study on Visual Exploration \\ of Model Simulations by Multiple Types of Experts}



\author{Nadia Boukhelifa\textsuperscript{1} \tab~Anastasia Bezerianos\textsuperscript{2} \tab~Ioan Cristian Trelea\textsuperscript{1} \tab~Nathalie Mejean Perrot\textsuperscript{1} \tab~Evelyne Lutton\textsuperscript{1}}
\affiliation{
  \institution{\textsuperscript{1}{UMR GMPA, AgroParisTech, INRA, Univ. Paris-Saclay}}
  \institution{\textsuperscript{2}{Univ. Paris-Sud \& CNRS (LRI), Inria, Univ. Paris-Saclay}}
  \institution{\{firstname.lastname\} \tab~{\textsuperscript{1}{@inra.fr} \tab~{\textsuperscript{2}}@lri.fr}}
}






\renewcommand{\shortauthors}{N. Boukhelifa et al.}

\begin{abstract}
Experts in different domains rely increasingly on simulation models of complex processes to reach insights, make decisions, and plan future projects. These models are often used to study possible trade-offs, as experts try to optimise multiple conflicting objectives in a single investigation. Understanding all the model intricacies, however, is challenging for a single domain expert. We propose a simple approach to support multiple experts when exploring complex model results. First, we reduce the model exploration space, then present the results on a shared interactive surface, in the form of a scatterplot matrix and linked views. To explore how multiple experts analyse trade-offs using this setup, we carried out an observational study focusing on the link between expertise and insight generation during the analysis process. Our results reveal the different exploration strategies and multi-storyline approaches that domain experts adopt during trade-off analysis, and inform our \hrem{design }recommendations for \hrem{SPLOM-based trade-off analysis tools }\add{collaborative model exploration systems}.  
\vspace{-5pt}
\end{abstract}



\begin{teaserfigure}
\vspace{-15pt}
\centering
  \includegraphics[trim=0.5cm 8cm 0.5cm 12cm, clip, width=0.85\textwidth]{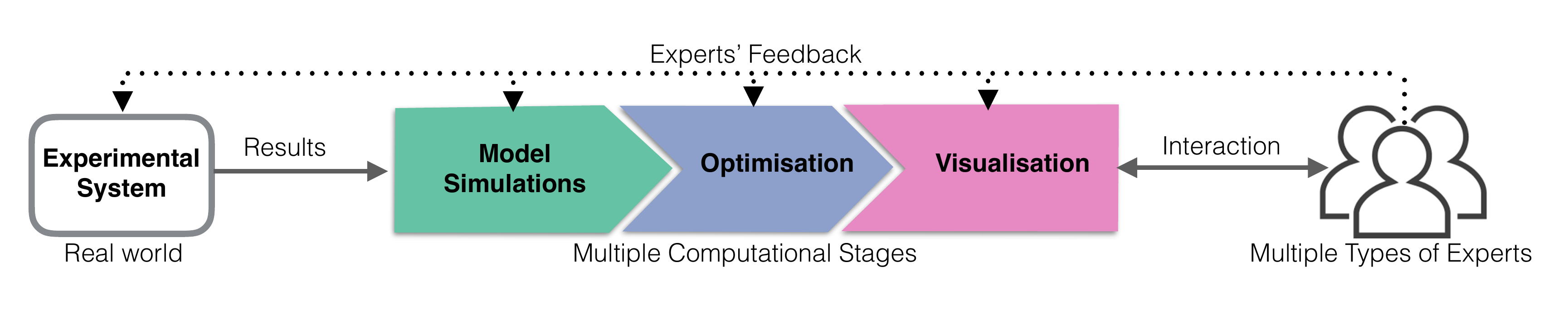} 
  \vspace{-20pt}
  \caption{The proposed model exploration framework: multiple experts interact with a visual representation of optimised simulation results. Experts' feedback impacts the exploration at all stages of the computational pipeline.}
  \label{fig:method}
  \vspace{5pt}
\end{teaserfigure}
\maketitle
\vspace{0pt}

\section{Introduction}

As our knowledge of complex processes\rmed{(biological, engineering, physical)} increases, to conduct their work domain experts rely often on the use of \emph{models}, i.e., abstract representations of entities and relationships within a specific domain or process. 
They have to manipulate these sometimes complex models for their daily work to reach insights, make decisions, and plan future projects.
%
For example, agronomic engineers who want to propose\rmed{precise, but} robust wheat fertilisation strategies to farmers, need to account for the wheat growth process and how it is affected by soil and weather conditions. 
To do this, they manipulate existing wheat growth models
~\cite{brisson1998,jeuffroy1999} 
that help them explore how, for example, a late fertilisation impacts wheat yield and quality, and whether the outcome changes depending on weather conditions. 
\rmed{In the agronomy domain, a number of established wheat growth models exist that can enable this investigation(e.g.,~\cite{brisson1998,jeuffroy1999}). However, }
To effectively manipulate such models, domain experts currently face three key challenges pertaining to: the \emph{multiple competing objectives} experts have to handle, the \emph{complex exploration space} they need to navigate through, and the \emph{multiple expertise} required for model understanding.

First, \addf{domain} experts have to deal with \emph{trade-offs}. Their work often attempts to 
reconcile multiple competing objectives in a single investigation~\cite{sedlmair2014}. For instance, agronomic engineers look for fertalisation strategies that on the one hand maximise yield, and on the other hand reduce the amount of supplied fertilisers, and the nitrogen loss\rmed{to the environment}
~\cite{ravier2018,cui2010}. 

%
 Second, \addf{domain} experts often resort to model simulations to explore trade-offs 
that 
can 
generate complex outcomes.
\rmed{However, the}
The results of such simulations are huge numerical data files which are \emph{difficult to explore} manually. Yet, finding robust 
solutions cannot be done automatically, since resolving conflicts between multiple criteria necessitates subjective human judgement~\cite{sedlmair2014}. For example, agronomic engineers use their expertise to specify the acceptable threshold for wheat quality\rmed{(e.g., in terms of percentage of protein content)}, and how much delay farmers would tolerate before they can supply fertilisers.

Third, understanding the model intricacies is challenging, even for domain experts.
Such models are often written by third parties \addf{(modelling experts or other domain experts)},\rmed{ are complex,} and require extensive technical knowledge to understand.
\rmed{Domain experts also have to \emph{align} their mental models of 
the domain\rmed{domain},
with numerical model assumptions and predictions.}
This challenge can be amplified when \addf{domain} experts are only specialists in part of the underlying process. 
For example, 
one agronomic 
engineer may be an expert in fertilisation strategies for\rmed{a local} one environment, but does not fully understand the impact of climate on other geographical regions. 
Given the \emph{complexity} of \rmed{both} 
the modelled process, the model intricacies and the exploration space, it is rare for a single \addf{domain} expert to \emph{fully} comprehend all those aspects in depth. 

Previous work with domain experts from agronomy~\cite{boukhelifa2017interactive}, 
highlights the lack of accessible setups and visual support tools to manipulate complex models and trade-off spaces. This observation may well apply to other domains where complex data and model spaces are analysed~\cite{boukhelifa2017workers}\rmed{ by non-professional data scientists 
~\cite{boukhelifa2017workers}}.\rmed{ In particular, there is need for 
support tools to enable model-based investigation by multiple domain experts, without having to go through the steep learning curve of writing a model or fully understanding an existing one.} 
\rmed{Our long term goal is to build such\rmed{ visual support} tools, but before doing that}
Before building such\rmed{ visual} support tools, 
we need to \emph{understand how domain experts explore models, including their trade-off spaces}.


We propose a simple setup for model exploration, and a user study that adopts it, in order to better understand how multiple domain experts explore complex models. Our user study is based on two model use-cases~\cite{mouret2015,jeuffroy1999}\rmed{, one for a wheat-crop model~\cite{jeuffroy1999}, and another for a wine fermentation model~\cite{mouret2015}}. Each use-case consists of two case studies of real scenarios explored by 
experts.\rmed{The study focuses on how multiple experts perform trade-off analysis tasks during model exploration, and on the links between insights and expertise.}
Our method (\autoref{fig:method}) consists of first conducting a Multi-objective Optimisation (MO) 
to reduce the model exploration space. This results in a multi-dimensional \emph{Pareto Front}, which is explored visually using a scatterplot matrix (SPLOM) and linked 2D views\rmed{presented} 
 on a shared interactive surface (\autoref{fig:study_setup}). In each use-case, we recruited participants with \addf{different 
 expertise}, covering the study domain\rmed{(biological processes)}, the numerical model, the simulation and optimisation, and the Pareto front visualization. Such a setup 
 ensures\rmed{was carried out to ensure}
 that the various types of insights and expertise are aligned, and 
 facilitates 
 the validation of the findings from different perspectives~\cite{Chuang2012}. 

We contribute: \rmed{The contributions of this paper are the following:} 
(1) A simple setup that facilitates the visual exploration and validation of complex models by synchronising expertise from the study domain, modelling, optimisation and visualization.
(2) An observational study using the setup, that inspects the role of expertise in insight generation during model-exploration and trade-off analysis.  \add{Our results revealed iterative analysis approaches with branching scenarios.} 
\hrem{(2) An observational study using the setup, that inspects the role of expertise in insight generation during model-exploration and trade-off analysis, and that revealed different exploration strategies and multi-storyline approaches adopted by the experts. }
\add{We identified analysis scenarios where \addf{multiple types of} experts examine together new and refined research questions and hypotheses (\emph{new}, \emph{refine}, \emph{alternative}, \emph{compare}), and other scenarios where \addf{they} \hrem{experts} learn to appropriate the tool and setup (\emph{initial}), and attempt to recap and establish common ground (\emph{storytelling})}. And (3) design recommendations for \hrem{SPLOM-based trade-off analysis} \add{collaborative model exploration} systems. 





\section{Related Work}


\add{The nature of sensemaking activities and the cognitive processes involved have been the subject of established work~\cite{pirolli2005,klein2007}, with recent studies focusing on how groups establish common ground~\cite{clark1991grounding,convertino2008,willett2011,gergle2013}, uncover hidden knowledge~\cite{convertino2009,goyal2016}, and engage with large analyses~\cite{feinberg2017,kery2018,passi2017,passi:2018,pine2015}. In terms of general sensemaking, the data-frame model \cite{klein2007} describes several key macrocogintive processes relevant to our own investigation, including connecting data to a frame (an explanation reflecting person's compiled experience), reframing, elaborating, questioning and comparing frames. The more recent work detail additional behavioural and analytical processes observed in collaborative settings.} 
\add{Our work is orthogonal, focusing on the role of expertise. }
\rmed{Research has looked at \emph{insight} and \emph{expertise} in relation to human performance and interactive systems. }\add{In terms of analysis methods, }our work is similar to ~\cite{guo2016} who performed an insight-based user study to understand how analysts reach insight during visual exploration.\rmed{ We contribute to this body of work a user study that examines the } Our focus, however, is on collaborative model exploration, in particular for trade-off analysis.

\subsection{Insight and Expertise}
Insight is considered the goal of visualization~\cite{card1999}. However, there is no consensus on the exact meaning of the term. North~\cite{north2006} focuses on the key \emph{characteristics} of insight, which are complexity, depth, quality, unexpectedness and relevance. \fix{In contrast, }Chang et al.~\cite{chang2009} distinguish spontaneous insight from the insight traditionally described in visualization, which they describe is about \emph{knowledge-building} and \emph{model-confirmation}, and where schematic structures, such as a mental model, are important to find patterns as well as to infer them. \fix{Furthermore,} Pousman et al.~\cite{pousman2007} explore the notion of multiple \emph{types} of insights for a broad range of user groups and describe analytic, awareness, social and reflective insights. Considering insight as a \emph{process} rather than the end result, Yi et al.~\cite{yi2008} describe four key processes of how people reach insight: provide overview, adjust, detect pattern, and match mental model. 

\rmed{In cognitive sciences, the spontaneous nature of insight discovery, or the \emph{aha-moment}, is often emphasised~\cite{bowden2005}. Insight is described as \quotes{the process by which a problem solver suddenly moves from a state of not knowing how to solve a problem to a state of knowing how to solve it}~\cite{mai2004}.} 
\rmed{Kunios et al.~\cite{kounios2014} describe other characteristics of insight such as the emotional response, and breaking from an impasse or a mental block. They stress that insight can be defined differently depending on which combination of features one selects.} 
\rmed{Insightful problem solving and deliberate analytical reasoning may be associated with different patterns of (resting-state) brain activity~\cite{kounios2014}. }

We follow Saraiya et al.'s~\cite{saraiya2005insight} definition of insight as \quotes{an individual observation about the data by the participant, a unit of discovery}.\rmed{This type of insight is more analytical, according to Pousman's et al.~\cite{pousman2007}: \quotes{analytic insights come from exploratory analysis, extrapolation, and consist in the large or small eureka moments where a body of data comes into focus for a user}.} The insights discussed in this paper are the result of \emph{confronting} single or multiple expertise with data and model artefacts. In terms of key characteristics, similar to Kounios et al.~\cite{kounios2014}, we consider insight any deep realisation, whether it happens suddenly or not.

Another type of knowledge that we are interested in in this study is \emph{expertise}. Expertise is broadly defined as highly-specialised domain-specific knowledge~\cite{chi2014nature}. 
\rmed{For visualization and human-computer interaction, expertise relates to human performance in areas such as tasks analysis, learning and training, interface design, and cognitive modelling~\cite{farrington2006}.} 
It can be in the form of \emph{tacit} knowledge, defined as knowledge that cannot be explicitly stated or transferred to other people~\cite{polanyi2009}; or it can be\rmed{There exists a different type of expertise,} more~\emph{declarative} or \emph{procedural}. This type of expertise can be articulated linguistically~\cite{farrington2006} and, thus, may be captured using think-aloud protocols, interviews and questionnaires. 

Glaser and Chi~\cite{chi2014nature} describe key characteristics of experts across domains. For instance,  
experts\rmed{ across domains} 
typically take a long time analysing a problem qualitatively before attempting to solve it. During this \emph{incubation period}, they try to comprehend the situation, build a mental presentation of its core elements, and review the problem from different angles before attempting to implement a solution~\cite{paige1966,skovholt2004}.


We are interested in insight generation during \add{collaborative model exploration }
and the role of explicit expertise in this context. We consider expertise \hrem{as }the background knowledge that allows domain experts to reach new and deeper insights from interacting with the model representation. \rmed{One can hypothesise that the more knowledge experts have in a 
domain, the more insights they can potentially generate. However, Fisher et al.~\cite{fisher2016curse} found that~\quotes{expertise can increase confidence in the ability to explain a wide variety of phenomena}, leading to what is known in the literature as the \emph{curse of expertise}. This occurs when \quotes{more knowledge leads to mis-calibrated explanatory insight}. We aim to calibrate this phenomena by involving multiple experts in a single exploration sessions.}

\subsection{Model Exploration}
Model exploration is an iterative process of discovery and refinement, which consists of examining the entities and relationships that underpin models. It can be carried out by the model builders themselves (\emph{model producers}), or by the problem owners who use such models to study a specific domain or process (\emph{model consumers}).  The exploration can be manual, automatic or semi-automatic.  Manual exploration consists of experimenting with different 
input parameters and any constraints to launch model simulations. 
Often this is a trial and error process~\cite{sedlmair2014}.
 Automatic exploration relies on algorithms, such as from genetic or evolutionary computation, to systematically explore the model search space. Semi-automatic approaches take into account human feedback to steer the exploration. 
The outcome of model exploration is often \emph{insight} that informs model design and implementation, or improves our understanding of the modelled phenomena. 

\rmed{Most model simulation experiments 
focus on\rmed{revolve around} 
parameter space exploration~\cite{reuillon2013,sedlmair2014}, where the goal is to \emph{understand} the link between model input and output. Understanding the parameter space is important to \emph{test} and \emph{validate} model predictions against experts intuitions or a known ground-truth. 
It ultimately enables users to \emph{steer} the 
computational processes towards desirable areas of the search space~\cite{mulder1998}.}

\rmed{In a recent survey, Sedlmair et al.~\cite{sedlmair2014} classified strategies to explore model parameter spaces into four main categories: informed trial and error, local-to-global, global-to-local, and steering. Furthermore, through a structured analysis of the visualization literature, they compiled a list of six exploration tasks: (1) \emph{optimisation}: to find the best parameter combination given some objectives; (2) \emph{partitioning}: to discover the different types of possible model behaviours (3) \emph{fitting}: to determine input parameters that best describe some measured data; (4) \emph{outliers}: to detect special output cases; (5) \emph{uncertainty}: to judge the reliability of model output; and (6) \emph{sensitivity}: to test the effect of input changes on the output.}


Our aim is to \emph{understand} how \rmed{domain} experts explore model parameter spaces in the presence of conflicting criteria.\rmed{(or objectives).} \rmed{We focus on user-model interactions and the exploration strategies that multiple types of experts adopt to manage trade-offs between competing objectives.}\rmed{Moreover, we 
study exploration strategies for an already optimised search space, rather than the exploration of the full model parameter space. }To the best of our knowledge, there are no user studies documenting how experts explore simulation model trade-off spaces using a collaborative setup that unites model consumers \addf{(study domain experts)} and producers \addf{(modelling experts)}.


\subsection{Trade-off Analysis \add{and Pareto Optimality}}
We call trade-off analysis the type of analysis users perform\hrem{, with the aid of interactive tools, \nb{Correction, not just with interactive tools}} to reach optimal decisions in the presence of multiple conflicting model objectives. \add{Conflicting objectives, or criteria, are typical when exploring alternative options in many domains including agronomy. Cost is typically one of the main criteria (e.g., financial and environmental cost of supplying fertilisers), together with some measures of desired quantity or quality (e.g., wheat yield and protein content).} \add{An approach to identify \quotes{good} solutions to multiple objective problems is \hrem{with }Pareto optimality~\cite{LopezJaimes2009,kung1975}. This technique produces a diverse set of compromise points between conflicting objectives. 
Within these points, there exists a set of non-dominated points called the Pareto Front (PF)~\cite{kung1975}, where no objective can be improved without sacrificing at least one other objective. \add{During trade-off analysis,} decision makers \hrem{ As such, it is up to humans to }select which PF solutions work best according to their expertise and preferences}.

Visualizing \add{and exploring the PF is an important step in multi-criteria decision making~\cite{tuvsar2015}. However,} representing this set for more than three objectives\hrem{ has long been a significant challenge for MO problems} is challenging~\cite{tuvsar2015,ibrahim2016,boukhelifa2018}\rmed{,boukhelifa2018}.
\rmed{A number of visualization techniques exist to present Pareto sets to humans, where the goal is to facilitate the exploration of alternative solutions and assist decision making. }
Tu{\v{s}}ar and Filipi{\v{c}}~\cite{tuvsar2015} provide a comprehensive review of visualization techniques for Pareto front approximations in evolutionary MO, including among others scatterplot matrices, \rmed{bubble charts, radial coordinate plots, }
parallel coordinates, \rmed{heatmaps, Sammon mapping, self organising maps,} 
principal component analysis, \rmed{isomaps,}
and their own technique called prosection. 

The scatterplot matrix technique shows all the bivariate projections of the solution space, presented in a table. This is considered to be an intuitive approach~\cite{tuvsar2015}, albeit not very scalable since decision makers cannot see all dominance relationships at once. In our study we consider small multi-objective problems, of up to eight objectives, and we use a large display to overcome the problem of screen space. 


\section{Exploration Approach and Setup}
\label{exploration-approach}

To support 
multiple experts explore complex models, we 
combine: 
data pre-processing, a visualization tool running on a physical setup\rmed{ that supports collaboration}, 
and a coverage of multiple expertise.
\vspace{-3pt}

\paragraph{Data Preprocessing.} The types of complex models we are dealing with\rmed{(e.g., the agronomy model of wheat fertilisation)} are often explored through simulations.
\rmed{Domain experts run multiple versions of these simulations using different input parameters, often numerical (e.g., 
geographical region, temporal window in the process to introduce the fertiliser, amount of fertiliser)
Depending on the complexity of the model, the number of input and output parameters can be large (often exceeding 20-30 dimensions in the cases of experts involved in our study).}Depending on their research questions, experts may run hundreds or even thousands such simulations, by altering the input parameters, resulting in possibly thousands of alternative simulation results\rmed{(data points)} that they wish to explore. 
\rmed{For instance, after specifying the input parameters of interest and their ranges, 
one wheat agronomic engineer run over 300k simulations. In the case of wine fermentation, a multi-objective algorithm sampled the search space over 1k\rmed{\nb{Cristian, correct? OK by CT.}} iterations.} Visualising the relationships between input and output parameters could help experts make sense of their data, but visually exploring such large datasets can be daunting. 

In an attempt to make the problem more tractable and easier to visualise, we rely on the observation that experts are often interested in reconciling multiple competing objectives~\cite{sedlmair2014}\rmed{ (\emph{trade-offs})}. 
\rmed{An important aspect of their work is to understand and reconcile multiple competing objectives~\cite{sedlmair2014}, often in a single exploration. 
For instance, our agronomic engineer searches for fertilisation strategies that maximise yield and reduce the amount of supplied fertilisers\rmed{(two competing goals)}~\cite{ravier2018,cui2010}.} 
\rmed{Based on this observation,}We adopt a pre-processing step of calculating the Pareto front~\cite{kung1975} of the simulation results, i.e., identify the list of \emph{non-dominated} data points (simulations) that are possible solutions to the \rmed{multi-objective optimisation} MO problem.\rmed{To compute this set of optimal solutions we require (i) a list of criteria to optimise (e.g., fertiliser amount, crop yield), and (ii) and optimisation algorithm to compute the Pareto front.} 
The criteria to optimise are specified \emph{a-priori} by the domain experts. To generate the Pareto front data, we used the state-of-the-art algorithm in MO optimisation NSGA-II~\cite{deb2002} when we had access to the model source code (wine use-case). In the absence of the latter, we used a classical MO algorithm~\cite{woodruff2013} as a simple filter applied to the simulation files data (wheat use-case).

\rmed{
Before the exploration session, 
an optimisation expert held a brief interview with domain and model experts to discuss the objectives they want to optimise and the parameters they want to explore. We explain in the user study details of our contact with experts prior to the exploration session. 
We recommended 
they 
limit their optimisation objectives and parameters to around 20. In practice this range is roughly the limit of both what multi-objective optimisation algorithms can efficiently handle~\cite{ishibuchi2008}, and what our SPLOM system can visualise. 
We collected their criteria and simulation results prior to the 
sessions. 
%
We note that the definition of what criteria to optimise may be an iterative process.} 

\rmed{
The next part is to run the optimisation algorithm to compute the Pareto front. The best MO optimisation algorithm to use is not a straightforward decision. Depending on, for example, the preference articulation stage (a-priori, a-posteriori or no preference articulation), and the availability of model source code and computational resources, 
different algorithms are more appropriate~\cite{marler2004}~\nb{cristian: is this correct?}\nb{CT= "What are preferences in this case? Domain experts specified  decision variables and their bounds, objectives, constraintS. OK?"}. 
We used the state-of-the-art algorithm in MO optimisation NSGA-II~\cite{deb2002} when we had direct access to the model source code (wine model use-case)\rmed{and domain experts to specify their optimisation preferences}. In the absence of the source code, we used a classical multi-objective optimisation algorithm~\cite{woodruff2013} as a simple filter applied to the simulation files data (wheat model use-case).}
\vspace{-3pt}

\paragraph{Visualization Platform and Physical Setup}
To visualise the Pareto front data points\rmed{(simulations)}, we used a 
SPOLM-based tool~\cite{elmqvist2008rolling,Boukhelifa:2013} that has been successfully used to explore multi-dimensional data. Scatterplot visulizations are often used to show relationships between two dimensions 
or characterise distributions of data points~\cite{rensink2010,dimara2018},
and thus are appropriate for visualising relationships between objectives (to show trade-offs). 
Our tool has a SPLOM, and allows multiple query selections (differentiated by colour) to help experts narrow their search space to important parameters 
~\cite{tuvsar2015}. The system also provides a \add{bookmark} history of past query selections and a means to store \quotes{favourite} views and queries. 
Finally, 
experts can 
enter new combined dimensions manually through a mathematical formula field, or to evolve them automatically. Previous work has shown that allowing experts to test new dimensions is important to the exploration of model trade-offs \cite{Boukhelifa:2013}. 
Given that we are targeting a joint exploration between multiple experts, we\rmed{ wanted to} provide a physical setup that could foster collaboration \cite{Isenberg2013}. As such, we instrument a room with a shared touch-enabled display that host the main visualization.\rmed{, and that could be seen by all experts.}

\paragraph{Expertise Diversity and Stages.}

From our experience in the domains of agronomy and food process engineering, there are three types of expertise that appear at \emph{multiple stages} in research related to complex processes~\cite{perrot2016}\rmed{perrot2011,}.  Domain expertise\rmed{(e.g., expertise on the biological process)} usually comes first and informs the next stage, the modelling process. Modelling expertise comes next, and communication between experts of these two stages can be tight, especially when models are first developed. It is also common for optimisation expertise to follow, for example, for parameter estimation and tunning~\cite{motta2013}. Optimisation experts often have to work closely with model builders and domain experts to understand model input and output, and to specify pertinent optimisation criteria and constraints. 
\rmed{What is important to note is that}
This three-way collaboration is \emph{iterative}, and currently appears to be carried out in an ad-hoc asynchronous manner. Importantly, a single expert rarely understands all stages\rmed{. In our use-cases for instance, only one participant had expertise in the domain, the model and the optimisation } (\autoref{tab-participants}). 

Our approach interjects another stage of expertise, that of interactive visualization.\rmed{, to facilitate collaboration between the various experts, and to aid the exploration of model trade-off spaces.} By bringing the four expertise together in a \emph{synchronous} fashion, we hope that our experts\rmed{, through the use of a visual analysis system,} will not only gain insights that can generate new research on the biological processes\rmed{(feeding back to the domain expertise)}, but also help form hypotheses about the modelling, optimisation and visualization processes. \rmed{Furthermore,}
\rmed{By confronting the various expertise \rmed{involved in the exploration use-cases,}
we hope to ensure that any insights reached in the exploration are validated by all expertise perspectives~\cite{Chuang:2012}.}

\rmed{Recruitment of such a diverse group of expertise can appear challenging. Nevertheless, we found through our interviews with experts during recruitment, that they themselves often know who the complementary experts are, and they are often already in contact (e.g., the optimisation expert with the wine fermentation domain experts). }

\begin{figure}
  \includegraphics[width=.95\linewidth]{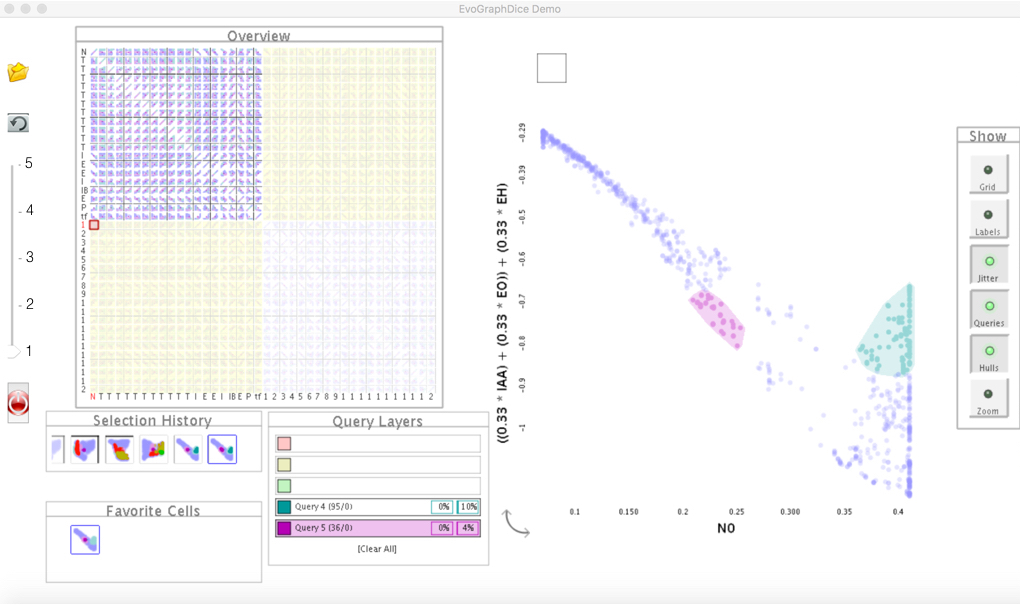}
  \includegraphics[width=.95\linewidth]{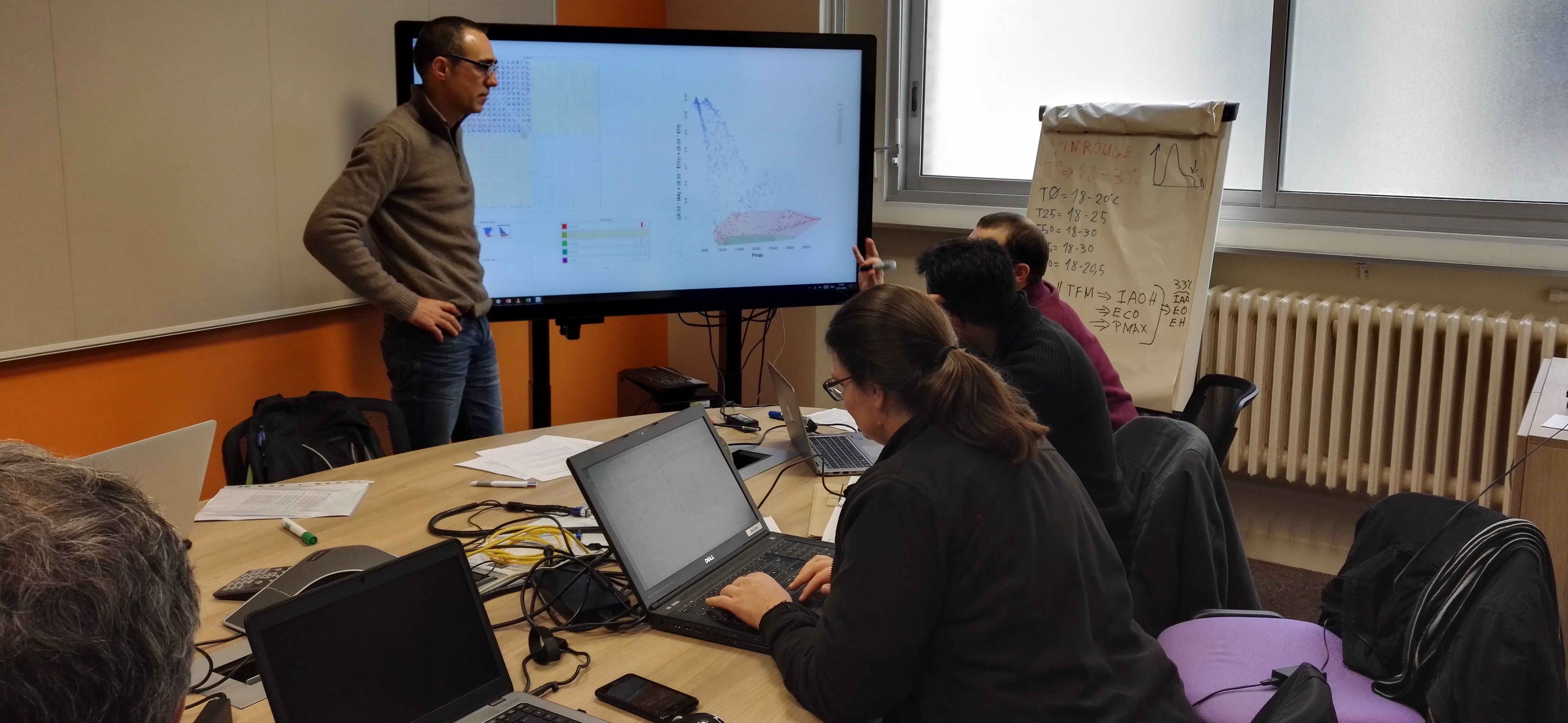}
  \vspace{-6pt}
  \caption{Study setup. (Top) \hrem{close-up of the shared display with the}\add{Screenshot of the SPLOM tool\hrem{running} with the matrix on the left and the main scatterplot on the right}. (Bottom): broad\hrem{er} view of the room, showing individual laptops and flipchart.
   Figure \hrem{captured during} \add{relates to }\hrem{the first} case study \hrem{of the wine model }CS1a.}
  \label{fig:study_setup} 
  \vspace{-10pt}
\end{figure}

\section{Study Design}
\label{sec-design}
We conducted an observational study to understand how multiple types of experts explore models using the SPLOM-based visualization tool described in~\autoref{exploration-approach}, when they are considering multi-criteria trade-offs. The goal of the study was to investigate the following research questions:

\begin{list}{[Q\arabic{rcounter}]}{\usecounter{rcounter}}
    \item what \emph{strategies} domain experts adopt to explore trade-offs during model exploration. 
    \item how \emph{insight} is reached during trade-off analysis.
    \item how domain experts \emph{interpret} findings from the visualization through model abstraction and optimisation.
    \item what role the different \emph{expertise} play in this context.
\end{list}

\subsection{Participants}

\paragraph{Expertise.} \rmed{In total, }We had 12 participants (five female) including four co-authors of this paper (optimisation and visualization experts). The mean age was 44.5 years.\rmed{ Seven participants self-identified as experts in the problem domain.
\rmed{microbiological engineering, bio/process and food engineering, physicochemistry, oenology, and agronomy.}} 
\rmed{Six 
were experts in the model we 
used to generate the corresponding Pareto front, five were optimisation experts, and four were experts in the visualization tool.} In terms of \emph{expertise overlap}, most participants had double expertise (8 participants), in particular domain-model expertise (5 participants) due to long term usage of the model \stat{mean 4.6 years}. One participant had three types of expertise domain-model-optimisation (\autoref{tab-participants}).
{Domain expertise ranged from 3--30 years \stat{mean 14.4}, 1--14 years for the models \stat{mean 4.2}, and 2--10 years for optimisation \stat{mean 5.4}. Apart from the visualization and optimisation experts (co-authors of this paper), only one domain expert had prior experience \fix{with }
the tool used in the study.}
All were researchers with titles such as lecturer, professor, \add{permanent} researcher or research engineer. One participant was a final year PhD student. 

\vspace{-5pt}
\paragraph{Recruitment.} Participants were\rmed{all} recruited from an agronomy research centre in \add{INRA} and collaborating institutions. 
The recruitment procedure drew on previous collaborations between the authors and the different participants, or between participants. However, these collaborations had not previously united all of the expertise covered in this study. 

\begin{table}[]
\centering \Small
\setlength{\tabcolsep}{2.0pt}
\begin{tabular}{llccccc} \hline 
\textbf{P\#} & \textbf{Field}              	& \textbf{UC} & \textbf{Domain} & \textbf{Model} & \textbf{Opti} & \textbf{Vis} \\\hline
1           & Microbiological engineering   		& 1                      & \checkmark      & \checkmark     &                       &         \\
2           & Bioprocess engineering   		 	& 1                      & \checkmark      & \checkmark     & \checkmark            &         \\
3           & Physicochemistry        		& 1                      &                 &                &                       &         \\
4           & Microbiological engineering 			& 1                      & \checkmark      & \checkmark     &                       &       \\
5           & Bioprocess engineering      			& 1                      & \checkmark      &                &                       &       \\
6           & Oenology                    	& 1                      & \checkmark      & \checkmark     &                       &       \\
7           & Process engineering        			& 1                      &                 & \checkmark     & \checkmark            &       \\
8           & Agronomy                    	& 2                      & \checkmark      & \checkmark     &                       &       \\
9           & Food engineering          			& 2                      & \checkmark      &                &                       & \checkmark  \\
10          & Optimisation                	& 1, 2                   &                 &                & \checkmark            & \checkmark  \\
11          & Optimisation                	& 1, 2                   &                 &                & \checkmark            & \checkmark  \\
12          & Visualization               	& 1, 2                   &                 &                & \checkmark            & \checkmark \\\hline                     
\end{tabular}
\caption{Study participants for the wine\rmed{fermentation} use-case UC1 and wheat\rmed{fertalisation} use-case UC2, their field of work, and expertise in the study domain, the model, the optimisation procedure and the\rmed{SPLOM} visualization tool used in the study.}
\label{tab-participants}
\vspace{-8mm}
\end{table}

\subsection{Study Procedure and Apparatus}

Prior to the study, an optimisation expert held a brief 
interview with domain and model experts 
to discuss the objectives they want to optimise and the parameters they would like to explore. They then 
calculated the Pareto fronts. 
The actual study was carried out in two sessions, that we call \emph{use-cases} based on the research area of the domain and model experts. In the first \emph{use-case}, a wine fermentation model was explored, and in the second a wheat-crop model. Each use-case was run in two parts (a and b), on two different days, that we refer to as \emph{case studies}\rmed{(e.g. CS1a for the first part of the first use-case).}. Each part lasted on average\rmed{2 hours and 3} 123 minutes, and a different Pareto front for the same model was explored. We made sure all types of expertise were present. 
There was no overlap of participants between the two use-cases, other than the visualization and optimisation expertise.  

\rmed{We used the same study setup for all exploration sessions. }Participants were seated around a\rmed{shared} touch-enabled surface, and\rmed{ a drawing board and} a flip-chart was available in case they wanted to take any\rmed{ hand-written} notes outside the tool. \rmed{In addition, p}Participants were asked to bring their own laptops, where we installed the same exploration software in order for them to perform a training task\rmed{, but also to use them for personal exploration if needed} (\autoref{fig:study_setup}). The study took place in one research institute hosting a 3M Multi-Touch, 65" UHD-4K display.\rmed{The visualization software run on a Windows machine with Intel(R) Core(TM) i-5 processor (3.5 GHZ, 16 GB RAM).} 

At the beginning of each\rmed{domain} use-case we\rmed{ first} introduced the tool, then we asked participants to perform a \textbf{training task}~\cite{Boukhelifa:2013}.\rmed{ This task was formulated as a game 
\cite{Boukhelifa:2013}. 
The goal was to ensure that all our participants were able to operate the visualization tool}\rmed{, and make use of its main functionalities}\rmed{ including query creation, brushing and linking, and the aggregation of dimensions.} The next part of the study consisted of an \textbf{open exploration task} using a Pareto front dataset that was prepared beforehand. Participants were encouraged to \emph{re-discover} what they knew about their data before \emph{searching} for new insights~\cite{Boukhelifa:2013}.

\rmed{Throughout the four case studies, }We asked participants to think aloud and verbalise their thoughts. A study facilitator\rmed{ (at least one author)} was present to answer experts' questions and discuss their findings. In some occasions, the facilitator asked participants to explain the insights they found and 
to clarify whether these were new or confirmations of known findings. The sessions were video recorded and user interactions with the visualization tool were logged. After the study,\rmed{all} participants filled in a short questionnaire to collect demographic information\rmed{ and self-assessment of their expertise for the domain, the model, the optimisation and the visualization. Expertise}, and to self-assess their own expertise (using a 5-point Likert scale, and number of years in practice). We considered participants experts if they selected \quotes{3} or above as their expertise score.\rmed{ Only one expert gave ratings lower than 3 for all expertise types.}


\subsection{Models and Pareto Fronts Datasets}
In both use-cases, we used an optimisation algorithm to create a Pareto front that combines objectives (dimensions to optimise) and other parameters$^1$\footnote{1. In the visualization tool both objectives and parameters are simple data dimensions.}~(\autoref{tab-models-pfs}). Domain experts in the first use-case sought to maximise wine aroma (three esters), and minimise undesired compounds (higher alcohols) and the energy required to control the fermentation in red (CS1a) and white (CS1b) wine. 
In the first case study of the wheat use-case (CS2a), domain experts searched for interesting wheat crop fertilisation strategies that maximise yield, protein content and minimise nitrogen loss to the environment. A second Pareto front was generated in the last case study (CS2b) as domain experts wanted this time to find fertilisation strategies that maximise yield and minimise nitrogen dose and loss. We note that the results of CS2a helped create the new Pareto front for the subsequent case study (CS2b), as domain experts identified new research questions that take into account a new set of objectives.

\begin{table}[]
\centering \Small
\setlength{\tabcolsep}{2.0pt}
\begin{tabular}{llccc}\hline
\textbf{Case Study} & \textbf{Model} & \textbf{\#Objectives} & \textbf{\#Parameters} & \textbf{\#Points} \\\hline
CS1a & Wine (red)~\cite{mouret2015} & 8 & 6 & 1000 \\
CS1b & Wine (white)~\cite{mouret2015} & 8 & 6 & 1000 \\
CS2a & Wheat~\cite{jeuffroy1999}& 3 & 15 & 64 \\
CS2b & Wheat~\cite{jeuffroy1999} & 3 & 19 & 13601 \\\hline
\end{tabular}
\caption{Models and Pareto fronts datasets.\rmed{ used in the study, per case study session.}}
\label{tab-models-pfs}
\vspace{-8mm}
\end{table}


\vspace{-3pt}
\subsection{Data Collection and Video Coding Events Identified} 
\label{sec:data-collection}
The data we collected consists of 494 minutes of video recordings,\rmed{automatically generate} log files
, and\rmed{ the } observational notes\rmed{ we took during the exploration sessions}. Video annotation was carried out using an annotation software (\emph{chronoviz.com}) in two main \emph{iterative} passes. In the \textbf{first pass}, two authors independently watched one hour of\rmed{ the first case study video} CS1a, and using an initial coding scheme, coded the video as a sequence of discrete events. 
The two evaluators then met to compare codes and resolve conflicts and inconsistencies.\rmed{ This resulted in the refinement of the initial annotation scheme and the addition of further categories and objects as described below}  
In the \textbf{second phase},\rmed{one of the initial coders and another} two evaluator\rmed{ then} used this scheme to annotate the rest of the videos.\rmed{Each time, one coder annotated the whole video, and a second annotated a calibration window (from 20 to 40 minutes). This time frame was selected based on the observation that the initial 20 minutes of each case study video tends to focus on how the Pareto front was generated and on tool exploration method, rather than on tasks directly related to trade-off analysis.} After coding each video, the two evaluators met to discuss annotation codes and resolve any new conflicts.\rmed{ This process was repeated for the remaining three case study videos.} The inter-rater reliability scores for the four case studies video coding were: 77.19\%, 88.63\%, 86.36\%, and 87.09\%. \rmed{respectively. \ab{An alternative measure is Cohen's kappa, to do only if time ...}}

\paragraph{\textbf{Events.}} A coded \emph{event} is\rmed{defined as an interesting observation with regards to our study research questions (Q1--4),}
 characterised by a \emph{start time}, \emph{type}, \emph{actor}, and \emph{object}. 
\rmed{We decided not to code event duration as this is hard to specify accurately, especially in a collaborative context.} The \emph{type} of the event was assigned from an initial annotation scheme related to our research questions, and 
agreed upon prior to coding. It had three coarse categories: Human-Human Interaction, Insight, and Expertise. 
An event is further characterised by an \emph{actor}, referring to the type of expertise used by the participant who triggered the event: i.e. domain, model, optimisation or visualization expertise (\autoref{tab-participants}). 
By event \emph{object}, we mean the point of focus of the event. Each event is assigned to a single category type, and is associated with a single actor and object. 

The final annotation scheme contains six \emph{main categories}:

\begin{list}{\labelitemi}{\leftmargin=0cm}
\item \textbf{Human-Human Interaction (HHI), 31.59\%}: 
\rmed{any }discussions between participants about\rmed{ any of } the event objects listed below. 
\item \textbf{Human-Computer Interaction (HCI), 41.32\%}:\rmed{ all} user interactions with the tool\rmed{ including SPLOM cell selections, lasso selections, and query creation. }\rmed{and interactions with the favourites album.}. We also coded whether a dimension is aggregated\rmed{(i.e. a combined dimension)} and whether it was an optimisation objective.
\item \textbf{Human-Paper Interaction (HPI), 2.56\%}:\rmed{describes } events where participants wrote down notes during the exploration\rmed{used the flip-chart or the board to write side notes during the exploration}.
\item \textbf{Insight, 13.84\%}: a surprising or unexpected finding about the objects below, 
\rmed{Similar to~\cite{saraiya2005insight}, the notion of insight also includes}including new research questions and hypothesis (11.35\% of all insights). We \rmed{further break this category into two parts:} also distinguish between \emph{new} (80.78\%) and \emph{update} (19.22\%) insights.\rmed{to denote new findings, and confirmed or updated insights respectively.}
\item \textbf{Expertise, 9.76\%}: 
explicitly expressed knowledge related to the domain, model, optimisation or visualization.
\item \textbf{Action, 0.9\%}: \rmed{action }plans that go beyond the current exploration, \rmed{, often as a consequence of new findings. For instance}e.g., plans to carry out a new biological experiments.\rmed{or to generate a new Pareto front.}
\end{list}

Three of the main categories (Human-Human Interaction, Insight and Expertise), operate on the following \emph{seven objects}:
\begin{list}{\labelitemi}{\leftmargin=0em}
\item \textbf{Alignment, 6.23\%}: \rmed{anything relating to }how well or not the observed patterns or findings relate to, or align with, existing knowledge in the domain, the model, the optimisation and visualization.
\item \textbf{Correlation, 11.44\%}: relationships or dependencies between data dimensions.
\item\textbf{Criteria, 21.09\%}: specification of data dimensions or combination of dimensions to use as a search criteria, to better answer a research question. \rmed{Experts use }Search criteria were created in the visualization tool using coloured lasso selections.
\item \textbf{Dimension, 6.45\%}:\rmed{ observations linked to the importance of a dimension, in general, with regards to the underlying phenomena or observed pattern.} events related to data dimensions, e.g.,\rmed{the weights of a dimension in a search criteria or a combined dimension, dimension range, and constraints on dimension.} weights of combined dimensions and threshold values. 
\item \textbf{Exploration Method, 19.57\%}: strategy used to explore the data or the model expressed in a non-technical way.
\item \textbf{Tool, 17.35\%}: technical details related to the 
 tool.
\item \textbf{Trajectory, 17.84\%}: exploration of the trajectory of one or more parameters/objectives, and their temporal evolution.\rmed{, and the interpretation of its meaning.} 
\end{list}

 

\paragraph{\textbf{Scenarios.}} While coding, we noticed that the exploration was structured in what appears to be\rmed{independent scenarios or } linked mini-stories, often with logical \emph{transitions}.\rmed{ We were interested in characterising those scenarios and their transitions.} In a \textbf{follow-up coding pass}, \rmed{two evaluators independently watched all case study videos, and } we marked the beginning of new scenarios, and their transition objects\rmed{Again, the coders met to discuss notes and resolved conflicts.} (\statnp{inter-rater reliability = 94.44\%}). 
We categorised the 70 identified scenarios into 
 six types:

\begin{list}{\labelitemi}{\leftmargin=0em}
\item \textbf{Initial, 4/70}: 
preliminary exploration, often with the aim to explain or verify the dimensions of the Pareto front data set, and to try out the\rmed{general} tool functionalities. 
\item \textbf{New, 17/70}: 
exploration that attempts to answer a new research question or hypothesis that is different from the previous scenario.
\item \textbf{Refine, 30/70}: 
refinement from a previous research question or hypothesis, e.g.,\rmed{ by adding or } modifying existing search criteria.
\item \textbf{Compare, 7/70}: 
\rmed{exploration that explicitly compares} contrasting current and previous scenarios, e.g., in terms of generated insights.
\item \textbf{Alternative, 3/70}: 
branching out from a previous research question or hypothesis to explore an alternative exploration path, e.g., focusing on different exploration objects.  
\item \textbf{Storytelling, 9/70}: 
re-cap scenarios to share findings, interpret results and summarise the exploration steps.
\end{list}

In total, we generated 3307 events, 1686 for the first use-case and 1621 for the second use-case.\rmed{In the scenario video coding pass, w} We identified 70 scenarios, 39 in the first use-case and 31 in the second. The mean duration of a scenario is $\approx$ 7 minutes (\statnp{min=1m}, \statnp{max= 25m}), and contains on average 47 events (\statnp{min=5}, \statnp{max=133}). 




\section{Analysis of Video Segments}
\label{sec:analysis}

Our analysis is based on 70 trade-off exploration scenarios we identified during data collection. We carried out \emph{three} types of analysis to answer our 
research questions, 
focusing on patterns and strategies of trade-off exploration [Q1], insight generation [Q2], and the role of expertise in this context, in particular for \fix{the} alignment of findings [Q3,Q4]. 
We also provide a qualitative overview of participants' subjective feedback on the exploration pipeline and the visualization tool.\rmed{use of our SPLOM-based visualization tool for trade-off analysis.} \rmed{Participant comments are translated from the original language, and are shown between quotes.}



\begin{figure*}[t!]
	\centering
    \begin{adjustbox}{minipage=\linewidth,scale=1.2}
  		\includegraphics[trim=0cm 15cm 5cm 0cm, clip, width=.5\linewidth]{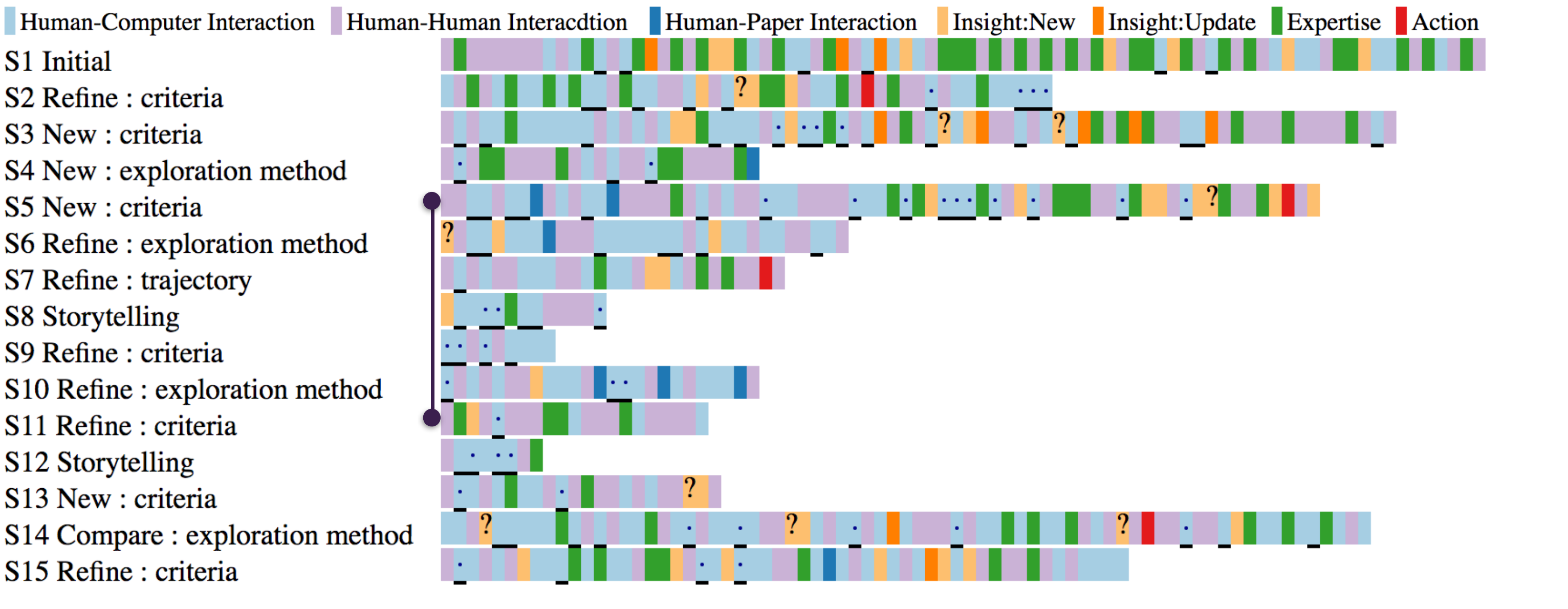}
  		\includegraphics[trim=0cm 15cm 5cm 0cm, clip, width=.5\linewidth]{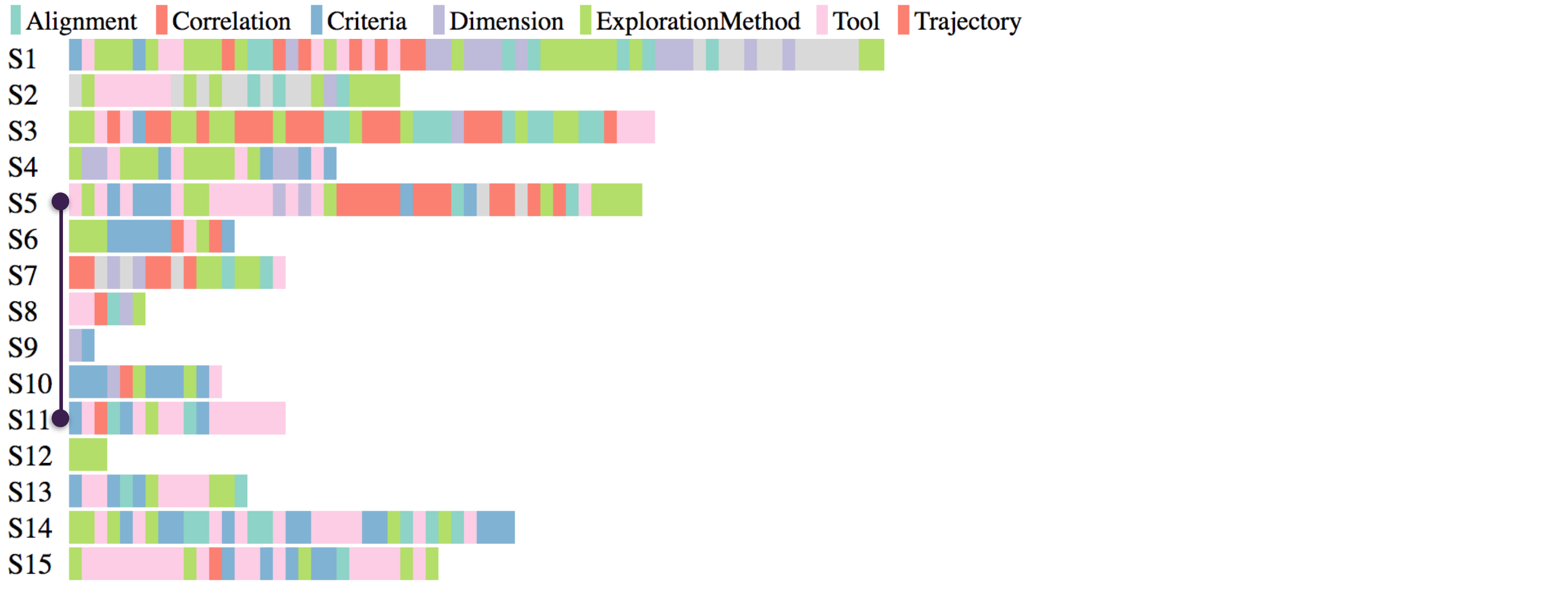}
  	\end{adjustbox}
  	\vspace*{-4mm}
  	\caption{\add{[left]} Scenarios of case study CS1a per category type. Chips\rmed{marked} with "$\cdot$" indicate \rmed{that the visualization included }views with combined dimensions, "$\_$"\rmed{indicates} for views with objectives, "$?$" for hypothesis \& research question insights.\rmed{Note that event chips do not indicate the duration of events. } Branching\rmed{ scenarios are} is indicated with a vertical\rmed{ black} line, e.g. between S5 and S11. \add{[right]} Scenarios of case study CS1a per object type. Complete scenarios can be found in supplementary material.  
	} 
  	\vspace*{-2mm}
  	\label{fig:scenarios_cat_o_view} 
\end{figure*}

\subsection{Scenarios and Trade-off exploration [Q1]} 
\subsubsection{Method}

\begin{figure}[h]
  \includegraphics[width=\linewidth]{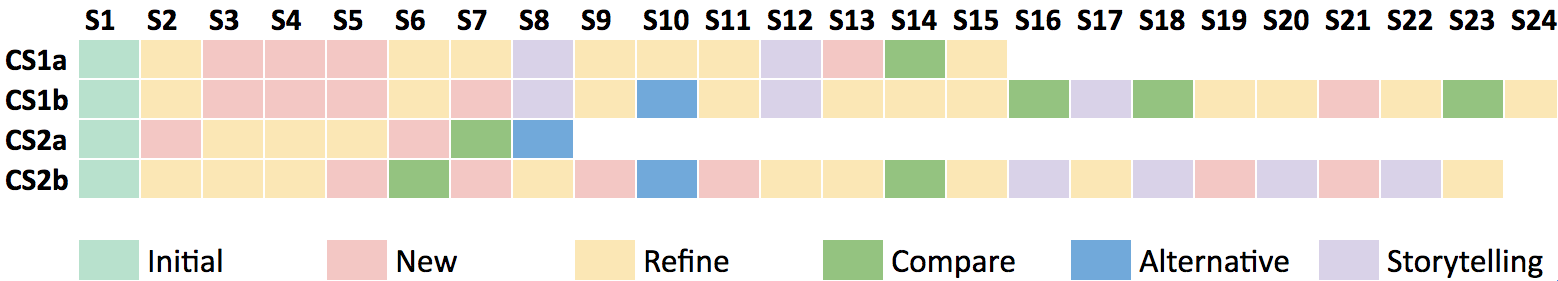}
  \vspace{-20pt}
  \caption{Trade-off analysis scenarios by case study: CS1a,b for the wine use-case, and CS2a,b for the wheat use-case.}
  \label{fig:sc_seq} 
  \vspace{-10pt}
\end{figure}

To analyse our results we took a top down approach. We first looked at scenarios and how they relate to the exploration. 
We created a visualization of \emph{scenario sequences} per case study session (\autoref{fig:sc_seq}), where each coloured chip corresponds to one identified scenario. 
The duration of the scenario is not encoded. To analyse the details of the scenarios themselves, we created two interactive visualizations of \emph{event sequences}. In the first visualization, coloured chips correspond to an event category type (\autoref{fig:scenarios_cat_o_view}-left), 
and in the second, to an exploration object type (\autoref{fig:scenarios_cat_o_view}-right). 
Each chip line (\emph{annotation line}), corresponds to one scenario. Scenarios are ordered chronologically, from top to bottom.

\rmed{For the \emph{categories annotation lines}, we colour-coded the six main event categories in~\autoref{sec:data-collection}.\rmed{,and we further split the insight category into: \emph{new} and \emph{update} insights.} 
By hovering over coloured chip in the annotation line, more information is revealed about the event including its object type and event start time. Furthermore, for the HCI event category, we indicate events operating on a combined aggregated dimension with a dot, and the objectives dimensions with a black underline. 
The \emph{objects annotation lines} provide a similar view on the data, but from the perspective of the explored objects. Hovering over chips in the object annotation lines reveals the main category of that event. Both visualizations allow filtering by event categories or objects.} 

\subsubsection{Overall Exploration and Collaboration Patterns}\label{sec:collab-patterns} 
\rmed{In this section we report on the overall exploration and collaboration patterns we observed from the visual analysis of the case study sessions (\autoref{fig:sc_seq}), the videos themselves and our observational notes.}
The visual analysis of scenarios in \autoref{fig:sc_seq} allowed us to make \emph{three} main observations on the exploration patterns: 

\indent $1.$ All study sessions start with an initial scenario where experts appropriate the visualization tool to explore their own datasets. 
\rmed{They tend to first}Here, they spend time analysing the problem qualitatively, building a mental map of the situation~\cite{chi2014nature}, before selecting an optimal solution from a 
set of alternatives.

\indent $2.$ We looked at frequency and order of scenario types. \emph{Refine} and \emph{New} are the most common  types of scenarios (\statnp{43\%} and \statnp{24\%} scenarios respectively), and they occur at any time during the exploration. 
 They both can follow as well as precede any other scenario type, except for the initial scenarios. This probably implies that the iterative nature of trade-off exploration is primarily due to the interleaving of \emph{new} and \emph{refined} hypotheses and research questions. 

It is interesting to note the difference between the two use-cases at the early stages of exploration (between $S1-S5$). There appears to be more \emph{exploration} for the first use-case illustrated by the contiguous blocks of \emph{new} scenarios (CS1a and CS1b), in contrast to the juxtaposed \emph{refine} scenarios in the second use-case (CS2a and CS2b), denoting more \emph{exploitation} behaviour. These two facets of exploration were also observed in previous exploratory visualization studies~\cite{Boukhelifa:2013}, and may correspond to how many research questions and directions participants had at the outset of the exploration. Additionally, the number of participants in the first use-case was higher, which might have led to the generation of many, perhaps diverging, hypotheses regarding the exploration objects. In the second use-case, there were less participants which may explain the more focused exploration.  


\rmed{\paragraph{3.}Most sessions (3 out of 4) finish with a refinement scenario. This is likely because experts are trying to drill down on some of their most important findings or insights. Looking more closely at the end of the exploration sessions, we noticed that most penultimate scenarios are a comparison (3 out of 4). This is likely because experts are trying to see if they can validate old insights or reach new ones using a different exploration method (CS1b-S14), or by following a new trajectory 
(CS1b-S23, CS2a-S7). 
More generally, comparison scenarios tend to occur towards the later stages of the exploration, often to shed a different light on the insights reached so far.\nb{Candidate for removal}.} 
\rmed{\paragraph{4.}}
%
\indent $3.$ The longer the exploration session, the more storytelling scenarios there are. Storytelling helps experts recap their findings before they continue their exploration. With more experts participating in the exploration (CS1a, CS1b), recaps are done periodically, presumably to \add{build common ground~\cite{carroll2006} and} ensure that everyone is on the same page. 
In smaller groups (CS2a, CS2b), recaps are concentrated at the end of the session.

We noticed that some scenarios reference earlier parts of the exploration, in a \emph{branch-out} fashion. As a result, we augmented the category and object annotation visualizations with \emph{jump-lines} linking the two scenarios in question.\rmed{This is the case, for instance in session CS1a-S11, which is a refinement of CS1a-S5 (\autoref{fig:scenarios_cat_o_view}-left). Initially, domain experts created a new search criteria based on a combined dimension they entered manually. In subsequent scenarios they refined this criteria and tuned their exploration method. In  CS1a-S11 they returned to the original search criteria to drill down to the individual data points behind the selection.} Back referencing scenarios\rmed{,or jump-lines,} exist in our dataset but they are hard to extract. We show in \autoref{fig:scenarios_cat_o_view}\rmed{ \autoref{fig:scenarios_cat_o_view}-right} one instance where this occurs, to illustrate the iterative and non-linear nature of trade-off exploration and sense-making~\cite{pirolli2005}.

When it comes to the setup, 
although we used a multi-touch interactive surface throughout the study, predominantly a single domain expert 
led most of the interaction in each case study session. The rest of the participants discussed seated or standing, but refrained from interacting. This finding is supported by other studies on collaboration around interactive displays~\cite{rogers2004,tong2017h}.\rmed{Furthermore, the fact that all participants did not have prior experience in exploring data on a large interactive surface, may explain why they adopted the traditional set-up of a presentation or a seminar.} We note that although most participants did not interact directly with the display, they were however actively involved in the exploration. For example, they proposed new research questions, requested to see particular views or to refine existing criteria. We also observed at least two instances in UC1 where domain experts explored the Pareto front in their own laptops.

\vspace{-4pt}
\subsubsection{\rmed{Varying }Complexity of Analysis Scenarios} 
The exploration scenarios identified vary in \emph{complexity}, which we characterise with two measures: number of events and scenario duration. 

In terms of average number of events per scenario type, \emph{Compare} scenarios had the most number of events \stats{74}{51}, 
followed by \emph{Alternative} \stats{68}{59}, 
\emph{Initial} \stats{59}{22}, 
\emph{New} \stats{47}{28}, 
\emph{Refine} \stats{44}{26}, 
and finally \emph{Storytelling} \stats{26}{32}. 
With the exception of \emph{Initial} and \emph{Storytelling} scenarios, overall the average time spent on each scenario type (in minutes) appears to be related to the average number of events in that scenario type: \emph{Compare} \stats{10.28m}{6.75}, 
\emph{Alternative} \stats{9m}{8.1}, 
\emph{Storytelling} \stats{7.33m}{8.15}, 
\emph{New} \stats{6.58m}{5}, 
\emph{Initial} \stats{6.25m}{3.59}, 
and \emph{Refine} \stats{5.9m}{3.74}. 
 Indeed we found that scenario duration and event count were strongly correlated \stat{Pearson's r(68) = 0.77, p $<$ 0.001}. Thus, in the next sections, to facilitate comparison between scenarios, we use scenario duration
 to normalise our results when reporting findings per and across scenario type (i.e., report frequency of events per minute).

\rmed{
\emph{Initial} and \emph{Storytelling} scenarios are particular. In early stages of exploration, there are many brief interactions with the tool in order to understand its functionality \stat{HCI=26.69\%}. Much of the discussions \stat{HHI=33.47\%} are centred around the tool and the exploration method (\statnp{53.16\%} of discussed objects), with little explicit examination of objects related to the problem domain such as trajectories \stat{5\%} or dimensions \stat{7\%}. During \emph{storytelling}, experts mostly discuss during these times \stat{HHI=47.08\%} and the exploration using the tool is limited to the replay of previous explorations \stat{HCI=31.66\%}.
}
We note that \emph{Compare} and \emph{Alternative} scenarios (10/70 scenarios), although rare, 
appear to be more demanding than the rest, based on our two complexity measures with on average more events and longer durations. 

\subsubsection{Search-Space Exploration Strategies}
\label{subsubsec:search-space}
Another way to characterise trade-off exploration strategies is to look at the types of scatterplots experts viewed. We delineate these strategies based on two orthogonal dimension properties: the \emph{granularity} of the projection axis for the viewed dimensions (single or aggregated), and dimension \emph{importance}. 
\rmed{(objectives and other parameters} 
\rmed{(primary or secondary).}\rmed{, depending on the priority experts assign to the criteria).} 
\rmed{Out of the 1367 HCI events in our annotation dataset, we coded all events that resulted in a change of the scatterplot view (i.e., selection of a SPLOM cell, the history manager or from the favourites album). 
These view change events correspond to \rmed{The scatterplot viewing data corresponds to} 
\statnp{54.13\%} of the total HCI events.} 

\vspace{4pt}
\indent $1.$
\emph{Dimension Granularity:} Although the SPLOM-based approach used in this study favours 2-by-2 viewing of dimensions,\rmed{experts used the aggregation functionality of our tool to help them explore their multi-dimensional Pareto fronts and combine dimensions simultaneously.\rmed{The scatterplot viewing data shows that }} experts consulted views having single dimension axes as much as they did for views having combined dimensions (\statnp{50\%} of views). However, the use of combined dimensions varied across the four case studies 
\stat{CS1a=11.62\%, CS1b=73.51\%, session CS2a=8.37\%, CS2b=6.48\%}. This is probably related to experts' research questions, and how far they reached in exploring the initial 2-by-2 search space. 
 In the first use-case, domain experts made extensive use of combined dimensions (in 34/39 scenarios), in particular in CS1b.\rmed{Aggregating made sense for the application domain,} Indeed, a wine fermentation recipe can be described as a linear combination of ratios of the various constituent aromas. 
In the second use-case,\rmed{experts were interested in exploring how a selection of optimal points (with respect to yield, nitrogen dose and loss) behave when looking at the details of a fertilisation trajectory (S8,19,29). 
Here, } experts preferred to explore fertalisation trajectories two steps at a time before aggregating.\rmed{, which they saw as something interesting: }
\rmed{\quotes{\emph{it helps to project more dimensions, not exactly in 3D, but on N with more dimensions...what would be interesting is to see at the same time when we select the dose we are going to supply, to see what iNN level we are covering.}}\Part{8}.} 

 
We note that 
experts rarely consulted views where both axes show combined dimensions (\statnp{2.97\%} of viewed plots).
This finding confirms results from a previous study~\cite{Boukhelifa:2013}, where domain experts prefer plotting aggregated dimensions against the original ones, to ground new insights. Moreover, experts never created nor consulted views with combined dimensions in the initial exploration stages, rather, they explored the original search space before attempting to aggregate.


\vspace{4pt}
\indent $2.$
\emph{Dimension Importance: } 
Most of the time, experts viewed scatterplots with at least one axis having one or more objectives (\statnp{63.10\%} of viewed plots). 
Overall, this indicates the importance of objectives in trade-off exploration, to make sure that subsequent data selections and filtering are optimal. However, this viewing rate varied per case study session 
\stat{CS1a=80.86\%, CS1b=99.31\%, CS2a=40.81\%, CS2b=12.36\%}.
%
The choice of consulted dimension type (objective or parameter)
likely varies depending on the research questions experts want to explore, and the chosen exploration strategy.

From our study notes and the storytelling scenarios, we observed that in the first use-case, experts predominantly used the objective space to \emph{drive}\rmed{ or guide their} the exploration. 
This strategy consists of first aggregating important objectives and then exploring the remaining dimensions in relation to that.\rmed{, hence the regular consultation of views having an objective dimension (denoted by an underscore line in \autoref{fig:scenarios_cat_o_view}-left).} An inverse strategy, used in the second use-case, consists of first finding 
ideal values for some parameters, 
and occasionally checking where this selection lies in the objectives space
, resulting in adjustments to the 
parameter 
values. 

In the presence of multiple objectives and parameters, experts created numerous criteria that they appear to \emph{prioritise}\rmed{and assign to them various degrees of importance.}. For instance, in the wine model use-case:\rmed{For example, the aroma criterion for the wine use-case consists of three objectives, and is considered more important than the cost criteria (another combined dimension of three energy objectives). 
During a storytelling scenario of the wine use-case, a domain expert reflected on the strategy they took to explore the data:}
\quotes{\emph{you start with the most important [aroma criteria], you select large, and then using your secondary criteria [cost criteria] you refine the selections}}) \Part{5}.\rmed{ For the wheat use-case, domain experts wanted to optimise the \quotes{loss} objective (primary criteria), more than they cared to always reach perfect yield (secondary criteria).}
\rmed{We note that data dimensions that are regarded more important, are often explored first. But the notion of importance may change across the different analysis scenarios, depending on participants' research questions and directions.} 
Regardless of whether the exploration is guided\rmed{primarily} by the objectives or by the parameters of the model, experts reached insights \stat{CS1a=56, CS1b=159, CS2a=87, CS2b=156 times} and aligned findings \stat{CS1a=37, CS1b=23, CS2a=23, CS2b=32 times}, indicating that both exploration strategies are valid.




\subsection{How Insight is Reached and Validated [Q2]}
We analysed how insight is reached during trade-off analysis, by looking at the scenarios and the objects of those insights, as well as the sequence of events that led to the insight. 

\vspace{4pt}
\indent
\emph{Insights by Scenario and Object Types: }
\rmed{Overall, i}Insights reached during the exploration sessions were mostly related to \emph{trajectories} \stat{42.57\%} and \emph{correlations} \stat{25.32\%}. This can be explained by the nature of the datasets explored in this study and the types of research questions our experts were interested in.
\rmed{\ab{check why?} Surprisingly, }\emph{Initial} scenarios generated more insights \stats{2.45}{1.92} (re-discovery of known findings), followed by \emph{Refine} \stats{1.59}{1.06} and \emph{Compare} \stats{1.56}{1.08}, then \emph{New} \stats{1.11}{0.53}, \emph{Alternative} \stats{1}{0.28} and finally \emph{Storytelling} \stats{0.34}{0.16}. 

\rmed{Due the nature of the modelled processes, a lot of the insights reached are about trajectories (e.g. \emph{Refine}:\statnp{56.06\%}, \emph{Compare}:\statnp{46.15\%}) }
\rmed{
The insights reached in \emph{Initial} scenarios are largely focused on correlations \stat{39.02\%} and trajectories \stat{24.39\%}, which are again tightly related to the research questions our experts set out to explore.\rmed{The correlation part was focused on finding and verifying relationships between primary and secondary criteria. The trajectories correspond to the wheat fertilisation  supplies over time, and sequences of temperature profiles for wine fermentation. None of those insights are related to the visualization tool since our participants were still discovering its functionalities.} \emph{Refine} and \emph{Compare} insights are also largely about trajectories (\statnp{56.06\%} and \statnp{46.15\%}\rmed{of the total number of insights respectively}). 
\rmed{For the remaining scenario types, insights were predominantly in the form of facts and findings about the various exploration objects. }For \emph{refine} and \emph{compare} scenarios, on average, most insights were about trajectories \stat{0.62 insight pm and 0.5 insight pm respectively}. 
Most insights of \emph{new} scenarios were about correlations \stat{0.37 insight pm}, and for \emph{alternative} scenarios, most insights were about criteria \stat{0.29 insight pm}. 
}
The few insights reached in the \emph{Storytelling} scenarios are mostly about the exploration method \stat{43.75\%}. This type of insight came from comparing methods experts previously used to analyse model simulation results (e.g., statistical analysis, experts decision rules), and the visual exploration proposed in this study. Not surprisingly, more than any other scenario type, a third of storytelling insights were in the form of a research question or a hypothesis\rmed{ to investigate in future explorations or in real biological experiments}\rmed{, based on the discussions 
that took place in the previous scenarios }. 
These types of insights were overall 
less common, and varied between \statnp{9.09\%} and \statnp{13.59\%} of insights per scenario type.

\rmed{All scenario types had predominantly more \emph{new} insights generated than \emph{update} ones (\emph{New} \stats{9.17}{7.61}; \emph{Update} \stats{2.15}{1.51}). 
This result is unexpected for the initial scenarios, as the study instructions encouraged participants to first verify what they already know in their data, before trying to find new insights. We note, however, that the relatively low rate of update insights could be due to the fact that our participants had not explored the Pareto front dataset prior to this study, and thus these indeed were new findings. However, it could also be due to a limitation of our study since it is difficult to separate new and update insights solely based on video sequences. In the absence of experts comments about the novelty of the finding, we coded such insights as new.}

\begin{figure}[t]
  \includegraphics[width=0.6\linewidth]{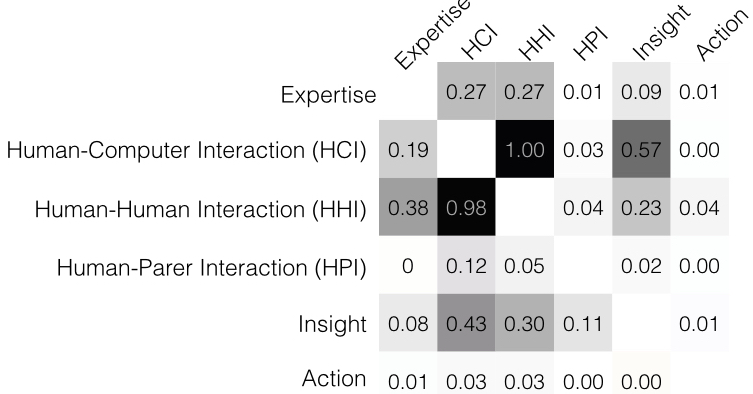}
    \vspace*{-4mm}
  \caption{A transition matrix of event sequences per exploration \emph{categories}. Events in rows precede events in columns, numbers and colour indicate frequency of transition. Here the most frequent sequence is  HCI-HHI. } 
  \label{fig:sequence_cat} 
  \vspace*{-5mm}
\end{figure}

\vspace{4pt}
\indent
\emph{Insights and Event Sequences: }
To extract event sequence patterns,\rmed{we made a first pass on the annotation data where} we removed repeated consecutive event categories from the annotation data.\rmed{ We did this per categories and exploration objects.} For instance, an event sequence HCI-HHI-HHI-HPI becomes HCI-HHI-HPI.\rmed{, Tool-Tool-Method becomes Tool-Method.} 
We then counted the occurrences of each \emph{two} subsequent events, and normalised each count by the maximum frequency value. The results are shown in the transition matrix in \autoref{fig:sequence_cat}\rmed{and \autoref{fig:sequence_object}}.\rmed{We note that our observations about the typical or most frequent event sequences do not imply a causality, but merely a temporal association.}

\rmed{
\begin{figure}[t]
  \includegraphics[width=0.7\linewidth]{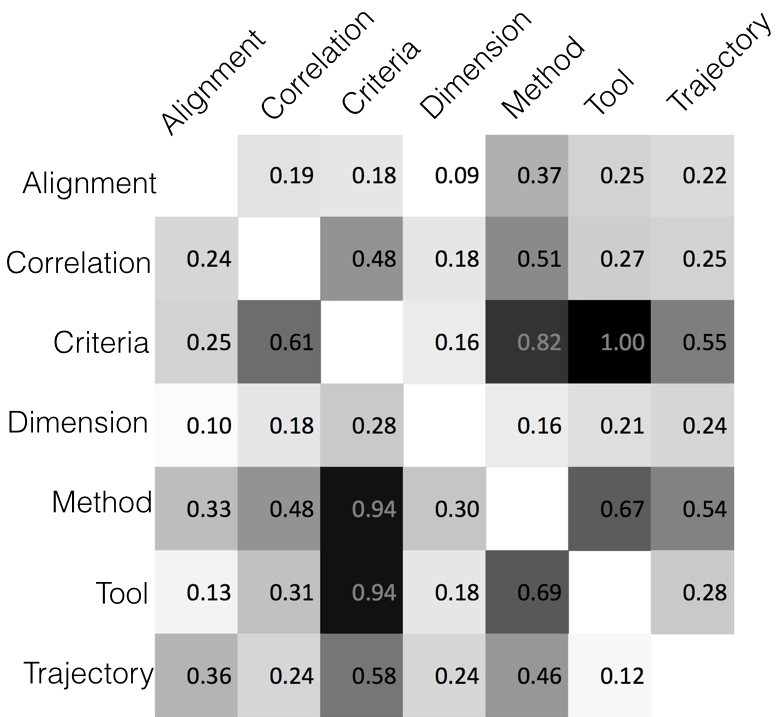}
  \caption{A transition matrix of event sequences per exploration objects. Events in rows precede events in columns. Here the most frequent sequence is Criteria-Tool.. 
  } 
  \label{fig:sequence_object} \nb{candidate for removal}
\end{figure}
}


\rmed{In terms of \emph{categories} (\autoref{fig:sequence_cat}), w}We found that the most frequent event sequences were HCI-HHI and HHI-HCI. Indeed, participants discussed their intentions, thoughts and findings as they interacted with our tool. Surprisingly, the least frequent sequences related to the \emph{insight} category (excluding Action and HPI) were Expertise-Insight (and vice versa). Rarely insight came just from articulating expertise. Rather, insight was most frequently reached after HCI or HHI events, whereas \emph{expertise} was most frequently followed by HCI or HHI (equally). 
\rmed{Moreover, insights can be followed by note taking. This can be seen visually in~\autoref{fig:scenarios_cat_o_view}-left. 
 Some detailed exploration (e.g. of trajectories) requires extensive note-taking during insight generation (e.g. to record parameters and their values).} \rmed{We can see this in CS1a-S18,20,21,22,24.}

\rmed{These findings may indicate the different nature of expertise and insight in the context of trade-off analysis.}

\rmed{
For the analysis of event sequences by exploration \emph{objects}, we found that the most frequent event patterns involve the following objects: \emph{criteria}, \emph{tool} and \emph{exploration method}. It appears that observations on these objects are closely related. For example, it may be that criteria selection influences the choice of exploration method, or that discussions on the exploration method often leads to experts exchanging on what search criteria to create in order to enable this exploration. More notably, discussions related to various search criteria (in the form of expertise, insight or HHI) are often followed by tool events, perhaps to discuss how to specify the criteria using the SPLOM-based visualization tool.\nb{candidate for removal + figure 5, sequences by objects.}
}

\rmed{
\emph{Trajectory} events are most frequently followed by criteria discussions. For example, at various occasions in the second use-case, domain experts started the exploration from an ideal trajectory, from which they made partial selections (i.e., search criteria). They then verified the correspondences between those selections and the remaining trajectory steps, as well as the objectives. \emph{Correlation} events are more frequently followed by discussions on an exploration method to validate the correlation, or by a search criteria, more likely to drill down to the details of that correlation. Finally, \emph{Alignment} events occur more frequently after discussions on the exploration method. Our experts often had to align what they found from the visualization, with what they knew from their domain. Often the discussions related to \emph{Alignment} covered the provenance of the Pareto front data.
\nb{candidate for removal.}
}

\subsection{Expertise [Q3,Q4]}
We illustrate through participants' quotes\footnote{\fix{All participants' quotes are translated from French.}} how expertise plays an important role during model exploration, particularly in: \rmed{. It is particularly important in } (i) identifying outliers and discrepancies, (ii) informing various types of decisions, and (iii) validating insights. 

\Part{8}, for instance, was able to quickly identify and explain an outlier, where wheat yield was unusually high: \quotes{\emph{I think it is a point for one year where there was a very favourable climate at the end of the cycle}}. On the other hand, she expressed uncertainty with respect to model predictions of a certain variable: 
\quotes{\emph{I do not trust the protein content level, wheat at 27\% of protein content, it is not something we see everyday!}}.

Domain expertise helped inform various types of decisions \rmed{related to dimensions, } such as which dimensions to aggregate in a search criteria and which to exclude: \quotes{\emph{then there is EAOH, and the famous higher alcohol which are precursors, which we need to have the least possible amount of, because it serves nothing from an aromatic point of view.}} \Part{5}. Similarly for \Part{8}: \quotes{\emph{in fact TP [protein content] does not need to be here [as objective], I just don't want it to be less than 11.5 ... for me personally losses [one objective] is more important than the other objectives}}.  
\rmed{Expertise can inform how participants decide to explore the model search space and the steps they take to do that: \quotes{\emph{in fact, in my head, since I have a notion of the trajectory of the [plant's] nutrition, I have this idea, it goes in the sense of the trajectory ... So, I start at N1.}} \Part{8}. The same domain expert also made suggestions to improve the optimisation procedure based on her expertise: \quotes{\emph{you have to be careful, such that you have at least one simulation with TP $>$ 11.5\% per year. Otherwise you will remove years, because TP is not at 11.5\% not because the fertilisation strategy was not good, but because the climate did not allow it}}.}

\rmed{The explicit knowledge provided by various experts facilitates the interpretation of insights and contributes to their validation.\rmed{For example, \Part{8} found a correlation insight linking protein content to year, which she was able to confirm based on her expertise: \quotes{\emph{Yes because maximising TP is dependent on the year}}. }}
A deeper validation of insights is reached when \emph{multiple expertise} are synchronised. Typically, observations in one part of the exploration framework, often the visualization, are aligned with existing expertise about the remaining steps of the exploration pipeline\rmed{ (optimisation, modelling and the study domain)}. For example, a correlation insight was found in the first use-case where two objectives (EH and EO) appear to be highly correlated. One optimisation expert raised a warning about this finding: \quotes{\emph{from the optimisation point of view they\rmed{ [EH and EO]} are correlated, but afterwards, we do not know}} \Part{7}. A modelling expert then confirmed that the observed correlation between the two objectives also exists in the model, and thus adding to the validity of this finding: \quotes{\emph{because these are the combinations I have found in order to reduce the model}}\Part{1}. Finally, a domain expert adds that this correlation also exists in the physical sense: \quotes{\emph{besides being correlated in the Pareto front and the model, EH and EO are also correlated in the metabolic sense}} \Part{5}. The same expert expressed the added value of the visualization in realising the nature of this correlation: \quotes{\emph{This [the correlation] we already have in the model, we see it experimentally, but here in the Pareto front it is a disaster\rmed{[too complex]} 
... You do not have a simple correlation between an alcohol and its ester}}\Part{5}. \rmed{A more general comment followed from the visualization expert advising participants on how to read the scatterplots given our exploration framework: \quotes{\emph{All that you know about the model, you will be able to find here. However, other things you find here [in the visualization], could also be interesting}}.}






\subsection{Subjective User Feedback}
\label{sec:subjective}

%
\indent
\emph{1. Reasoning about the world through multiple abstractions: }
Our participants liked the visual style of model exploration and how the framework helped generate complex new insights: \quotes{\emph{\rmed{the way I see it [the main result of the exploration], is that}we made a first exploration filter that made it possible to arrive at this decision tree, it's not negligible}} \Part{9}. Participants appreciated the ability to explore \emph{alternatives} within an already optimised space. For example, 
\rmed{Another finding, this time for the wine use-case,}the wine experts found two distinct fermentation strategies that led to the same aromatic composition through different temperature management strategies: \quotes{\emph{this is very interesting ..., my aim is not necessarily to reach the max of aromas,\rmed{say 0.75 of the different aromas,} either I use 0.4 which is my max\rmed{[for the initial nitrogen N0]}, or use 0.2-0.25. The result product is the same even though the nitrogen quantity is different}} \Part{5}.\rmed{ The same participant added: \quotes{\emph{it is a real super intelligent exploration ... this type of exploration is complicated, we cannot do it without EvoGraphDice [the tool]}}.}

However, there were some issues related to the interpretation of the visualizations\rmed{, expressed by the domain experts}: \quotes{\emph{it is strange that I do not have other points [for that year], ah no, because it is already optimised}} \Part{8}. Another domain expert added: \quotes{\emph{what is not clear from me is ... we reason in relation to what we know in terms of what is correlated to what\rmed{ [from experiments]}, but here in the Pareto front, there could be things that are very correlated, which may not be correlated [in reality or in the model]. This could be disturbing in terms of reasoning}} \Part{4}. This participant suggested overlaying the Pareto front points on the simulation data\rmed{(e.g. using a different colour)} as a reminder that these are optimised points. 

\rmed{Another issue was related to the non-assisted manner in which participants currently explored the trade-off space, which made the exploration sometimes cumbersome: \quotes{\emph{these decision rules\rmed{[values for the nitrogen fertalisation trajectory]}, it's too heavy! ... What disturbs me, having the same [fertalisation] trajectory for wheat with very low iNN level and one with correct iNN level, is very strange... it does not respect the physiology of the plant}} \Part{8}. This domain expert would have liked to more \emph{systematically} exploit the expertise that she previously formalised and tested, for example, to eliminate a number of non-viable fertalisation strategies.} 

\vspace{5pt}
\indent
\emph{2. Using SPLOM-based tools to explore model trade-offs: }
\rmed{Our participants liked the gradual selections and filtering of data across the 2D views: \quotes{\emph{for me, what I also like, is to be able to make chained selections [brushing and linking]}} \Part{8}. Their expertise was important in creating meaningful search criteria that helped them reach optimal solutions to their optimisation problems. In particular, they appreciated the tools' lasso selections, and the brushing and linking of views, which helped them prioritise their filters, and thus manage conflicts between the competing objectives: \quotes{\emph{what we did in terms of coloured selections, it is also a way to do the prioritisation of criteria}} \Part{5}. As described earlier (\autoref{subsubsec:search-space}), participants may prioritise dimensions differently depending on the analysis scenario, which can imply that any explicit support for chained filtering and prioritisation needs to be flexible and adapts to changing research questions and directions during the exploration.}
\rmed{In terms of missing tool functionalities that may have limited the trade-off exploration in this study, the most important ones were related to the search criteria and lasso selections, aggregation of dimensions, sensemaking of the exploration process, and the dynamic linking of the various stages of our exploration framework.} 
We identified \emph{four} main areas where the exploration tool may be improved based on participants comments and interactions. \textbf{(1)}\rmed{although multiple participants helped define each search criteria dimensions, as well as the size and location of the corresponding lasso selections, our tool does not directly support the collaborative editing of search criteria and lasso selections. Moreover, accuracy is an important factor to consider here, as our participants tended to start with large selections based on the primary dimensions, but then fine tune as they progressed in the exploration.} our tool lacked 
\rmed{support for fine touch-based lasso selections \rmed{on the tactile display}, and}
ways to import search criteria to other exploration sessions. \rmed{Moreover, for each use-case, participants explored multiple Pareto fronts, generated from the same models. Participants were interested in applying the same selections and search criteria to different Pareto fronts, but currently our visualization tool does not allow for exporting or importing work from one exploration session to another.}
\textbf{(2)} The text based dimension editor and detailed lasso selection were difficult to operate on the tactile surface.\rmed{, thus this event was the most time consuming of all HCI events. }
\textbf{(3)} Although a history of past selections and current queries are provided in our tool, participants often felt lost during the exploration\rmed{Particularly for longer analysis scenarios, participants struggled to make sense of their their coloured selections, and how they arrived to a particular view or search criteria }: \quotes{\emph{what was the basis of the reflection here? In fact, we seem to go faster than we have time to note down}}~\Part{6}. 
Our participants resorted to note-taking and storytelling to overcome this issue, but this was not sufficient. \textbf{(4)} The different components of our exploration framework are currently not linked dynamically. \rmed{Thus, if domain experts want to}Adding new objectives requires the optimisation procedure to be launched off-line, the results are then uploaded to the visualization tool.\rmed{Given the intensive calculations required to generate each Pareto front, this linking needs to consider the progressive viewing of intermediate results~\cite{zgraggen2017}.}

\vspace{-5pt}
\section{Discussion and Summary of Findings}
\label{sec:summary}
\add{To the best of our knowledge, there are no other studies that looked at collaborative model exploration in real-world settings. In terms of results, we found similar processes to those described in general sensemaking literature~\cite{klein2007,pirolli2005}. Our contribution here, however, is in identifying why these processes tend to occur (e.g., storytelling to recap), and when they occur (e.g., storytelling periodically for large groups, or at the end of big chunks of exploration for smaller groups). Alignment is a particular process to model exploration, also described in~\cite{Chuang2012}. However, our work considers more complex models, multiple computational stages and co-located expertise.} We discuss next \emph{seven} key findings from our study, relating them to our initial research questions, \add{and comment on the applicability of our methods to other domains}: 
 
\vspace{3pt}
\indent
\emph
{\add{1.} Exploration as Multiple Linked Analysis Scenarios} [Q1]
\label{sec:summary_scenarios}
We observed 
that\rmed{the patterns in how our study participants explored and collaborated. The}
exploration is split into\hrem{ several} 
mini-exploration scenarios. Trade-off analysis starts with a preliminary exploration often leading to a focused research question. The remainder of the exploration is characterised by the non-linear interleaving of new and refined hypotheses and research questions~\cite{pirolli2005}, operating on a variety of exploration objects. \add{Those scenarios denote a shift in the research questions and hypotheses set out by experts, which often result in change of focus in the model or data space. A parallel can be drawn between our approach and the data-frame model described in~\cite{klein2007}. For instance, their \quotes{reframing} maps to our new scenario, and \quotes{elaborating the frame} to our refine scenario. In our case, however, re-framing revealed itself to happen specifically when participants shift their research questions and hypotheses. Furthermore, we provide a more fine grained analysis of the exploration, by crossing high level categories of interest (e.g., insight, expertise), with exploration objects (correlation, exploration method).}

\vspace{3pt}
\indent
\emph
{\add{2.} Heterogeneous Analysis Scenarios} [Q1]
\label{sec:summary_scenarios_nature}
Analysis scenarios vary in length and complexity. Most of the exploration relates to refining research questions. Scenarios where experts actively compared or followed alternative exploration paths\rmed{to consider trade-offs,} were rare, tended to be longer and had more interaction events\rmed{ (as they delve into details about one scenario, or reach an impasse in another)}. 
While the absolute length of scenarios differs, their duration and event count were strongly correlated. 

\vspace{3pt}
\indent
\emph
{\add{3.} Storytelling \add{as a Grounding Process}}  [Q1,Q3]
\label{sec:summary_storytelling}
Participants engaged in the storytelling of their past exploration~\cite{riche2018data,tong2018storytelling},~\add{both in terms of gained knowledge, and the processes used to reveal the various insights~\cite{clark1991grounding,convertino2008}. This grounding phenomena helps groups with multiple expertise to establish a basis for working together across differences, and makes the sensemaking process through model and algorithmic abstractions more transparent~\cite{goyal2016,passi2017}}. 
\rmed{but at various rates that seem appropriate to their collaboration style and set-up. }{}
\fix{In our study}, storytelling scenarios are mostly discussion-based,
but happen likely for different purposes.  \rmed{happening periodically, but likely for different purposes.}
Large groups re-cap periodically to bring everyone up to speed. For smaller groups, storytelling tends to occur towards the end of the exploration, to summarise and prepare for the next session. 
\rmed{In both cases,  the longer the exploration session, the more the storytelling.}

\vspace{3pt}
\indent
\emph
{\add{4.} Two Key Strategies for Trade-off Analysis} [Q1]
\label{sec:summary_strategies}
Domain experts create multiple search criteria that they then prioritise. 
Objective dimensions are important to guide the exploration when used in the \emph{initial} search criteria, or to validate 
parameter criteria choices and selections a \emph{posteriori}. The first strategy seems more appropriate for model exploration scenarios where there are known or hypothesised connections between primary and secondary dimensions\rmed{(e.g. UC1 on the wine model)}, or when analysts are very familiar with the space, and so they are more likely to experiment. The second case seems more appropriate for under-explored models\rmed{(e.g. UC2 on the wheat model)} when experts are still trying to understand the model behaviour.\rmed{, which is likely to result in the creation of a new Pareto front.}  
In both cases, however, insight is reached, making both exploration strategies valid. 

\vspace{3pt}
\indent
\emph
{\add{5.} The Importance of Aggregating Dimensions} [Q1]
\label{sec:summary_aggregation}
Our analysis of the search space exploration strategies showed the importance of aggregating dimensions. 
SPLOM-based tools project data 
two dimensions at a time, whereas our participants were examining trade-offs between at least five objectives. The degree to which experts used dimension aggregation varied between 
sessions depending on 
their knowledge 
about the non-aggregated search space. 

\vspace{3pt}
\indent
\emph
{\add{6.} Multiple Pathways to Insight} [Q2]
\label{sec:summary_insight}
Our \rmed{analysis of how insight is generated shows that our} 
interactive model exploration framework allowed the generation of new insights and the confirmation of old findings. 
Similar to the tasks described by Yi et al.~\cite{yi2008}, our participants provided re-caps of their exploration (provide overview), continuously refined their exploration (adjust), discovered interesting relationships between data dimensions (detect pattern), and aligned different types of findings (mental model). We found that insight is reached in all scenario types including initial ones, and is often in the form of new surprising findings. In the context of our study, those insights were mostly about trajectories and correlations. An unexpected finding was that insights can also be reached during storytelling, 
mostly about expanding previous discussions and forming hypotheses and actions for future biological experiments. This highlights the importance of supporting storytelling in model exploration.\rmed{ We also found that scenario refinements can lead to new insights, which means that even \quotes{failed} scenarios (in terms of new insights) can be revisited and become more successful.} 

\vspace{3pt}
\indent
\emph
{\add{7.} The Indirect Link Between Expertise and Insight} [Q2,4]
\label{sec:summary_expertise_insight}
\rmed{During the exploration, our study participants demonstrated a deeper understanding of their respective domains, and appear to perceive a large amount of meaningful patterns or chunks of organised knowledge ~\cite{chi2014nature}. Their}Expertise appears to lead to insight and also to follow from it (e.g., to explain and validate findings), although in both cases not directly but with interleaving discussions and human-computer interaction events. Expertise was important in identifying outliers and discrepancies, informing various types of decisions (related to both analysis findings and exploration method), and validating insights. 
In our study, synchronising multiple expertise seems to provide deeper insights and validation of findings, and increase model understanding and confidence in the results, although more studies are needed to confirm our intuitions. 

\rmed{\subsubsection{Lack of Dedicated Tools for Trade-off Analysis}
Our participants appreciated the visual style of model exploration and how the framework helped them generate complex new insights. However, reasoning with an optimised parameter space was not very intuitive. Even though experts explored a reduced search space, they expressed a need for assistance to guide their exploration and navigation. 
Participants were quickly able to adopt the SPLOM-based visualization tool to perform trade-off analysis tasks, but it appears to lack key functionalities specifically tailored to Pareto front analysis tasks such as the ones described in~\cite{tuvsar2015}. 
In particular, there is need to support the creation and management of multiple search criteria, aggregation of dimensions, sensemaking of the exploration process, and the dynamic linking of the various stages of the exploration framework.\rmed{(e.g. interactive optimisation of model simulation results).}}

\vspace{-3pt}
\rmed{\subsubsection{\add{8.} Applicability \rmed{of our approach }to other domains}}
\subsubsection{\add{8.} Broader Applicability}
\label{sec:summary_gen}
\add{Our hypothesis-centred approach to characterise our exploration scenarios is well suited to how our domain experts conduct their research in agronomy. However, the approach can generalise to any domain with large input and output simulation spaces that considers trade-offs (e.g., finance, urban planning), and where the human investigation relies on what-if scenarios (e.g., intelligence analysis). Furthermore, our coded exploration objects (with the exception of \quotes{tool} that is SPLOM specific) apply to general data exploration with multiple expertise and can be a basis for future analysis. Finally, our model exploration framework can use any multi-dimensional visualization tool. \rmed{, even though the \quotes{tool} exploration objects we coded are specific to SPLOM exploration.}
}

\vspace{-8pt}

\section{Recommendations for Design}

Collaborative frameworks such as ours resembles participatory design approaches,\rmed{, for the purpose of model building, tuning and exploration.} where the presence of multiple expertise may improve the collective understanding of the modelled processes~\cite{boukhelifa2017workers}. 
 The importance of involving different types of expertise during model-based exploration for decision making, and having people who can speak the language of multiple experts, has also been documented~\cite{simon2015bridging}. Improving model understanding may have societal implications such as encouraging model-driven debate and discussions~\cite{von2010designing}. Our \add{study} findings\add{, and participants' subjective feedback (indicated below by \subjective{})}, suggest several ways that \add{collaborative} model exploration may be enhanced to better support \add{sensemaking during} trade-off analysis: 

\vspace{5pt}
\indent
\emph
{Methodology: } The think-aloud protocol used in this study allowed us to observe knowledge articulated by our participants, either as new findings (insight), or existing knowledge (expertise). Other methods for externalising or articulating \add{hidden} knowledge needs to be investigated~\cite{mccurdy2018framework}\hrem{, especially for tacit knowledge}\add{, particularly for collaborative settings}. \hrem{Also,}\add{The} externalised knowledge may then be exploited to augment the underlying data and models~\cite{baudrit2010towards}. \hrem{These methods need to distinguish between newly gained and existing knowledge.} \add{More generally, however, the distinction between expertise and insight}\hrem{Such a distinction} may inform the design of knowledge-assisted visual analytics tools, and can help refine insight-based evaluation of visualization systems. 



\vspace{5pt}
\indent
\emph
{\add{Making Sense of the Exploration Process: }}
\add{During the storytelling scenarios we identified, participants relied mostly on their own memory, the notes they took and the bookmark facility of our SPLOM tool, to reflect and share their own understanding of the main insights found \add{\textbf{(section 6.3)}}. However, they reported that this was not sufficient to make accurate reflections about their exploration processes (\textbf{\autoref{sec:subjective}\subjective{}}). 
Better support for sensemaking of the exploration process itself is needed, in particular, to reveal the hypothesis-driven approach experts took during their analyses \add{\textbf{(section 6.1)}}. This can be achieved by providing ways to visualise the exploration history and \fix{to} mark scenario stops and transitions, particularly for scenario comparison or alternatives \add{\textbf{(section 6.2)}}, and by allowing experts to record exploration values and the different types of insights reached \add{\textbf{(section 6.6)}}. 
}


\vspace{5pt}
\indent
\emph
{Trade-off Exploration Tools \& Setups: }\rmed{ Based on our findings (\autoref{sec:summary}), and participants' feedback indicated by (\subjective{}) below, }\rmed{ We make the following design recommendations for trade-off analysis tools:}We recommend that 
trade-off analysis tools should: 
\textbf{(a)} Support the creation of multiple search criteria, and help users keep track of their provenance and priority over time \add{(\textbf{section 6.4})}; 
\textbf{(b)} Allow\rmed{interactive} dimension aggregation\add{, and default combined dimensions, such as the results of principal component analysis \textbf{(section 6.5)}}; 
\textbf{(c)} Allow the dynamic regeneration of Pareto fronts\add{\textbf{\subjective{}}};
and \textbf{(d)} Allow multiple linked exploration setups, where people can create \emph{private} instances and explore alternatives individually, before sharing results with others \add{(\textbf{section 5.Q1})}.
\section{Study Limitations}
Our study focuses on biological applications, although we believe many of our findings generalise to other domains. We acknowledge \hrem{two }\add{four} main limitations of our study. \emph{First},\rmed{the 12 study participants formed two user groups corresponding to the wine and wheat use-cases. I} it is hard to realise a collaborative setup that unites multiple types of expertise, or to repeat the process in a distributed fashion. The two use-cases happened at eight-months interval because we needed to prepare the different stages of the pipeline. \rmed{(coordinate before hands on model, parameters of the multi-objective optimisation, the visualization set-up).}\rmed{In practice, the availability of multiple expertise at the same time was a blocking factor for having more groups for the study.}\rmed{Despite of this, the study observations and findings are rich.}\emph{Second}, although experts reached insights during their exploration\rmed{ of the model trade-off spaces}, it is hard to validate those findings from the biological sense. Those insights, however, allowed them\rmed{ to verify old findings, and} to prepare new research questions and future experiments.\rmed{ \emph{Third}, it has already been acknowledged that conducting an insight-based user study is both difficult and time-consuming. In particular, the complexity of video coding and insight validation grows with the number of participants~\cite{guo2016}.} 
\add{\emph{Third}, given that we are working with real experts and their problems, we did not control for expertise. Our method does not rely on expertise overlap. However, we noticed that our experts had gained new types of expertise over time, e.g., a domain expert became model expert after working with the model over a number of years. In the absence of expertise overlap and previous collaborations, we expect longer exploration sessions to introduce the various expertise and establish common ground. \emph{Fourth}, in terms of screen size, the large screen facilitates the display of high-dimensional datasets for groups of experts. If a smaller size screen is used, the visualization tool would need to allow for workspace awareness, and may lead to different results.}


\section{Conclusions}
We presented a multi-stage model exploration framework that unites multiple expertise, and a user study that adopts it. The focus of the study was on model-trade-off exploration, and on the link between insight and expertise during trade-off analysis. Our findings highlight a rich multi-storyline approach experts adopt during exploration\add{, where they constructively combine diverse expertise to resolve conflicts between competing objectives, and reach new insights.} 
\add{More work is needed to better understand collective as well as individual experts' role in generating insight in the different analysis sub-scenarios identified in our work (such as compare or storytelling), and how the different experts interact with the tool. These studies can help improve our understanding \hrem{of the human and collaborative aspects of work with large datasets, as well as } of the role of human expertise and its interplay with visual analytics in building common ground and externalising hidden knowledge.} \hrem{More work is needed to create visual analytics tools that assist various types of experts when working with complex models.}



\bibliographystyle{ACM-Reference-Format}
\bibliography{FP_exploration}

\end{document}